\documentclass[a4paper, 11pt]{article}
\pdfoutput=1
\usepackage{jheppub}

\usepackage{graphicx}  
\usepackage{dcolumn}   
\usepackage{bm}        
\usepackage{amssymb}   
\usepackage{color} 

\usepackage{amsmath}

\usepackage{epstopdf}
\usepackage{bm}       
\usepackage{amsmath}
\usepackage{cancel}
\usepackage[normalem]{ulem}

\newcommand{\bitem}{\begin{itemize} }
\newcommand{\eitem}{\end{itemize} }
\newcommand{\benum}{\begin{enumerate} }
\newcommand{\eenum}{\end{enumerate} }

\newcommand{\be}{\ensuremath{\beta} }

\newcommand{\cN}{\ensuremath{\mathcal N} }
\newcommand{\cO}{\ensuremath{\mathcal O} }
\newcommand{\gc}{\ensuremath{g_c^2} }
\newcommand{\gGF}{\ensuremath{g_{\rm GF}^2} }
\newcommand{\gtc}{\ensuremath{\widetilde g_c^2} }
\newcommand{\gtGF}{\ensuremath{\widetilde g_{\rm GF}^2} }

\newcommand{\kcr}{\ensuremath{\kappa_{\rm cr}} }
\newcommand{\gsim}{\ensuremath{\gtrsim} }

\newcommand{\MSbar}{\ensuremath{\overline{\textrm{MS}} } }

\newcommand{\vev}[1]{\ensuremath{\left\langle #1 \right\rangle} }

\newcommand{\eq}[1]{eq.~(\ref{#1})}
\newcommand{\fig}[1]{figure~\ref{#1}}

\newcommand{\secref}[1]{section~\ref{#1}}
\newcommand{\refcite}[1]{ref.~\cite{#1}}

\newcommand\redout{\bgroup\markoverwith
{\textcolor{red}{\rule[.5ex]{2pt}{0.4pt}}}\ULon}
\title{The renormalization group step scaling function of the 2-flavor SU(3) sextet model}

\author{Anna~Hasenfratz$^1$}
\author{Yuzhi~Liu$^1$}
\author{Cynthia Yu-Han Huang$^2$}
\affiliation{$^1$Department of Physics, University of Colorado, Boulder, CO 80309, USA}
\affiliation{$^2$Institute of Physics, National Chiao-Tung University, Hsinchu 300, Taiwan}
\emailAdd{anna@eotvos.colorado.edu}
\emailAdd{yuzhi.liu@colorado.edu}
\emailAdd{cynthia622@gmail.com}

\abstract{
We investigate the  discrete $\beta$ function of the 2-flavor SU(3) sextet model using the finite volume gradient flow scheme. 
Our results, using clover improved nHYP smeared Wilson fermions, follow the (non-universal) 4-loop \MSbar perturbative predictions closely up to $g^2 \approx 5.5$, the strongest coupling reached in our simulation.  
At strong couplings the results are in tension with a recently published work using the same gradient flow renormalization scheme with staggered fermions. 
Since these calculations define the discrete $\beta$ function in the same continuum renormalization scheme, they should lead to the same continuum predictions, irrespective of the lattice fermion action.

In order to test systematic effects in our computation we compare two different lattice operators, three different flow definitions, and two volume extrapolations. We find agreement among these different approaches in the continuum limit when the gradient flow parameter $c\gtrsim0.35$. 
Considering the potential phenomenological impact of this model, it is important to understand the origin of the disagreement between our work and the staggered fermion results. 
}
\begin{document}
\maketitle
\flushbottom

\section{Introduction} 

The 2 flavor SU(3) sextet model has received considerable attention lately as   candidate to describe the Standard Model Higgs boson as a composite particle~\cite{Aad:2015yza,Aad:2015owa,Hong:2004td,Dietrich:2005jn,Dietrich:2006cm,Arbey:2015exa,Fodor:2012ni,Fodor:2012ty}. Lattice simulations show that the model has a light $0^{++}$ scalar state that is well separated from the rest of the spectrum~\cite{Fodor:2014pqa}. Quantum corrections could lower the scalar mass to match the Higgs boson, while the vector meson is predicted in the  ~2TeV range~\cite{Foadi:2012bb,Fodor:2015vwa}. If the model is chirally broken, it has 3 Goldstone bosons as needed for electroweak symmetry breaking. If it is conformal, an additional  4-fermion interaction might drive it into the chirally broken regime~\cite{Fukano:2010yv}.  In either case the 2-flavor sextet model  is a promising strongly coupled  system that deserves detailed non-perturbative investigations.

The infrared properties of the model  are not yet settled.  2-loop perturbation theory predicts an infrared fixed point (IRFP) around $g^2 \approx 10.58$. The 3- and 4-loop (non-universal) \MSbar predictions for the IRFP are lower, around $g^2\approx 6.28$ and 5.90 respectively, while the mass anomalous dimension in all cases is $\cO(1)$~\cite{Ryttov:2010iz}. In a series of lattice calculations that used different improved  Wilson fermions  the authors of \cite{Shamir:2008pb,DeGrand:2010na,DeGrand:2012yq} found that the renormalization group (RG)  $\beta$ function  of  the Schr\"odinger functional gauge coupling became smaller than  the 2-loop perturbative prediction for $g^2 \gsim 4.0$ and possibly developed a conformal IRFP around $g^2\approx 6.0$. Ref. \cite{DeGrand:2012yq} predicted the mass anomalous dimension of this  fixed point around $\gamma_m \lesssim 0.45$. However these early works were not able to take the proper continuum limit.

There are several recent studies that use staggered fermions. In order to describe 2 flavors the  fermions are rooted. Rooting is not expected to be a problem when the proper continuum limit $g_0^2\to 0$ is taken at a chirally broken system, though the procedure   is less understood in conformal or walking theories near an IRFP. Finite temperature studies could distinguish conformal and chirally broken behavior but, as the results of refs. \cite{Sinclair:2014cga,Kogut:2014kla,Kogut:2015zta}  demonstrate, the phase diagram of the  2-flavor sextet model is not easy to characterize. While the results of refs. \cite{Sinclair:2014cga,Kogut:2014kla} appear to be consistent with chiral symmetry breaking, the latest results reported in \refcite{Kogut:2015zta} favor the conformal scenario. The LatHC collaboration has been pursuing comprehensive, large-scale studies of this model   for several years. Their results, presented in  a  series of papers,   support chiral symmetry breaking~\cite{Fodor:2012ni,Fodor:2012ty,Fodor:2014pqa,Fodor:2015vwa}. In  \refcite{Fodor:2015zna} the collaboration investigated the step scaling function, and found that   for gauge couplings  $g^2 \lesssim 6.5$ it is monotonic and shows no indication of an IRFP. 

In this paper we describe our investigation of the step scaling function  using Wilson fermions.  We use the same gradient flow renormalized coupling scheme as \refcite{Fodor:2015zna} but find that the continuum limit extrapolated step scaling function significantly deviates from the 2-loop perturbative prediction in the $g^2 \approx 4.0 - 6.0$ region, where it follows quite closely the (non-universal) 4-loop \MSbar prediction. We consider various systematic effects that could influence our results. We compare gradient flow couplings defined with two different operators, the plaquette and clover forms. We also compare several different gradient flow schemes based on the t-shifted gradient flow coupling~\cite{Cheng:2014jba}. These gradient flow schemes  predict the same continuum limit but have different lattice artifacts at finite bare coupling.  We also vary the lattice volumes used in the continuum extrapolation. We find that if the gradient flow parameter $c\gtrsim 0.35$, different variations predict the same continuum extrapolated step scaling function.
 Considering the phenomenological importance of this system, it is imperative to understand and resolve the  more than $3\sigma$ tension between \refcite{Fodor:2015zna} and our results.

In the next  section we briefly describe the calculation of the step scaling function using the gradient flow method. In \secref{sec:numerical} we outline the lattice action used in this study,  discuss the phase diagram, and problems related to  the tuning of the fermion hopping parameter to the critical surface. Section \ref{sec:analysis} outlines the analysis that leads to our result while in  \secref{sec:cutoff} we discuss  cut-off effects. Section \ref{sec:conclusion} concludes this study.

\section{\label{sec:stepscaling} The step scaling function with  gradient flow coupling} 

The gradient flow is a continuous and invertible transformation~\cite{Narayanan:2006rf,Luscher:2009eq} that  
has been used in a wide variety of applications. (For recent reviews see~\cite{Luscher:2013vga,Ramos:2015dla}.) 
The renormalized  gradient flow coupling ~\cite{Luscher:2010iy} is defined at energy scale $\mu $ as
\begin{equation}
  \gGF(\mu) = \frac{1}{\cN} \vev{t^2 E(t)}\, , \label{eq:gGF}
\end{equation}
where  $E(t) = -\frac{1}{2}\mbox{ReTr}\left[G_{\mu\nu}(t) G^{\mu\nu}(t)\right]$ is the energy density after flow time $t$ and $\mu = 1 / \sqrt{8t}$. The normalization \cN is set by requiring that $\gGF(\mu)$ matches the continuum \MSbar coupling at tree level.

At finite bare coupling the gradient flow renormalized coupling \gGF has cut-off effects that depend on the lattice action,  the  gradient flow transformation, and  the lattice operator used in defining the energy density $E(t)$.
While  changing the lattice action is usually not practical, considering several  flow transformations and/or lattice operators helps to investigate and reduce   cut-off effects. In the present study we compare two lattice operators, the plaquette and the clover operators, to discretize the energy density~\cite{Luscher:2010iy}. We also consider several definitions of the gradient flow coupling based on the t-shift approach introduced in \refcite{Cheng:2014jba}. The t-shifted gradient flow coupling is defined as
\begin{equation}
  \label{eq:t-shift}
  \gtGF(\mu; a) = \gGF(\mu; a) \frac{\vev{E(t + \tau_0 a^2)}}{\vev{E(t)}},
\end{equation}
with $|\tau_0| \ll t / a^2$.
The  $\tau_0$ $t$-shift can be either positive or negative.
In the continuum limit $\tau_0 a^2 \to 0$ and $\gtGF(\mu) = \gGF(\mu)$.
At lowest order the effect of the $t$-shift can be described as an $\cO(a^2)$ correction to the coupling $\gtGF(\mu; a)$; therefore, by optimizing $\tau_0$ the leading $\cO(a^2)$ cut-off effects can be removed.
In previous studies of   the 4-, 8- and 12-flavor SU(3) systems~\cite{Cheng:2014jba,Hasenfratz:2014rna}\, the optimal $\tau_0$ depended only weakly on $\gtGF(\mu)$ and simply setting it to a constant value was sufficient to remove most observable lattice artifacts throughout the ranges of couplings explored in each case. Alternatively, by considering various values of the t-shift one can identify not only  the leading $\cO(a^2)$ but the remaining higher order cut-off corrections as well~\cite{Hasenfratz:2015xpa}.
Since the gradient flow is evaluated through numerical integration, replacing $\gGF \to \gtGF$ by shifting $t \to t + \tau_0$ can easily be incorporated  in the analysis, without recalculating the flow.

In step-scaling analyses one connects the energy scale to the lattice volume $L^4$ by fixing the ratio $c = \sqrt{8t} / L$, as described in Refs.~\cite{Fodor:2012td, Fodor:2012qh, Fritzsch:2013je}.
Different choices of $c$ define different renormalization schemes,  predicting  different discrete \be functions in the continuum limit.
If periodic boundary conditions (BCs) are used for the gauge fields, these \be functions are only one-loop universal~\cite{Fodor:2012td}.
Ref.~\cite{Fodor:2014cpa} proposed to use a  modified  renormalized coupling 
 \begin{equation}
  \label{eq:pert_g2}
  \gc(L) = \frac{1}{\cN} \frac{1}{C(L, c)} \vev{t^2 E(t)}
\end{equation}
to perturbatively correct for cutoff effects. Here the function $C(L, c)$ is a four-dimensional finite-volume sum in lattice perturbation theory, which depends on the gauge action, flow and operator. Its  tree level expression is given in \refcite{Fodor:2014cpa}.
Since we use periodic BCs for the gauge fields, the correction $C(L, c)$ also includes a term that accounts for the zero-mode contributions~\cite{Fodor:2012td}.

We combine the t-shifted coupling of \eq{eq:t-shift} with \gc of \eq{eq:pert_g2} and use  the resulting \gtc gradient flow running coupling to investigate the  discrete \be function corresponding to scale change $s$,
\begin{equation}
  \label{eq:beta}
  \be_s(\gtc; L) = \frac{\gtc(sL; a) - \gtc(L; a)}{\log(s^2)}.
\end{equation}
Our final results for the continuum discrete \be function $\be_s(\gtc) = \lim_{(a / L) \to 0} \be_s(\gtc, L)$ are then obtained by extrapolating $(a / L)$ to 0.
If $c = \sqrt{8t} / L$ is fixed, the different values of $\tau_0$  and different energy density operators should all predict the same $\be_s(\gtc)$ in the continuum limit. Any deviation signals cut-off effects that are not removed by the $(a / L) \to 0$ extrapolation. These systematical errors can be   controlled either by increasing the lattice volumes used in the continuum extrapolation  or by increasing the parameter $c$~\cite{Fritzsch:2013je}.


\section{\label{sec:numerical} Numerical details}
\subsection{Lattice action} 

Wilson fermions are often plagued by bulk first order phase transitions that prevent simulations to reach the strong coupling regime. Improving the action by smearing the gauge fields in the fermion determinant or by including additional gauge terms can help, as was illustrated in  refs.~\cite{Shamir:2008pb,DeGrand:2010na,DeGrand:2012yq},  whose authors investigated the step scaling function using the Schr\"odinger functional running coupling of   2-flavor sextet fermions.  Using thin link Wilson-clover fermion \refcite{Shamir:2008pb} could probe the system in the range of $g^2 \lesssim 2.4$. Improving the action by using nHYP smeared fermions in \refcite{DeGrand:2010na} extended the range to $g^2\lesssim 5.5$. Including a smeared plaquette term in the gauge action pushed the first order phase transition to even stronger coupling, where yet an other issue arises~\cite{DeGrand:2012yq,DeGrand:2012qa}. On rough gauge configurations  zero modes in the U(3) projection of the  fermion smearing transformation can occur, a known problem in projected smearing transformations like nHYP or HISQ. To solve this problem  \refcite{DeGrand:2014rwa}   added a pure gauge term suppressing the zero modes of the nHYP smearing   to the action. This dislocation suppressing term, $S_{\text{NDS}}$, significantly reduces the occurrence of zero modes  and allows simulations deep in the strong coupling regime.
  
We have tested several versions of the nHYP smeared clover-Wilson fermion action for this project.
We considered reduced nHYP fermions where the smearing coefficients  ($\alpha_1=0.5, \alpha_2=0.5,\alpha_3=0.4$) are chosen instead of the optimal original parameters  ($\alpha_1=0.75, \alpha_2=0.6,\alpha_3=0.3$).
While the reduced coefficients do not provide the same suppression of the ultraviolet fluctuations as the original ones, they  give significant improvement~\cite{Cheng:2011ic}. At the same time reduced nHYP smearing cannot have zero modes in the smearing projection, making numerical simulations stable even without the $S_{\text{NDS}}$ term. We tried reduced nHYP smeared fermions with a simple plaquette gauge action, with a gauge action that  included a smeared plaquette~\cite{DeGrand:2012yq},  and also tested a plaquette gauge action with a negative adjoint plaquette term~\cite{Cheng:2011ic}. While numerical simulations were stable, all three actions were limited by the occurrence of first order phase transitions that prevented simulations to reach  couplings $g^2 \gsim 4$. 

Adding an NDS term to  the action proved to be significantly better than any  other choices.
 In this work we use nHYP smeared fermions with original smearing coefficients  ($\alpha_1=0.75, \alpha_2=0.6,\alpha_3=0.3$),  plaquette gauge action and an NDS term with coefficient $\gamma=(\gamma_1=\gamma_2=\gamma_3=)0.075$.    Perturbatively  the NDS term shifts the gauge coupling as
   \begin{equation}
   \frac{1}{g^2_0} = \frac{\beta}{2N_c} + \frac{\gamma}{N_c}(\frac{1}{3}\alpha_1 +\alpha_2+\alpha_3)\,
   \end{equation}
   where $\beta$ is the coefficient of the plaquette gauge term. With our choice of parameters this shift amounts to
   \begin{equation}
   \frac{1}{g^2_0} = \frac{\beta+0.1725}{2N_c} ,
      \end{equation} 
   a fairly small effect.
We found that simulations with this action are stable even at $\beta=0.5$, though the step size of the molecular dynamics evolution has to be decreased significantly at strong bare couplings.

\subsection{Tuning to the critical surface} 

Since our main goal is to determine the step scaling function, we generated configurations on symmetric $L^4$ volumes with  $L/a=8 - 28$. We chose   standard boundary conditions (periodic for the gauge fields and periodic in space, antiperiodic in time for the fermions) and 
tuned the fermion hopping parameter $\kappa$ to the critical surface.  

The critical $\kappa_\text{cr}(\beta)$ line in the ($\beta$, $\kappa$) plane is defined through the vanishing of the quark mass $m_q$. At zero temperature, the chiral limit of the Wilson type fermion action is taken along this line. We define $m_q$ through the axial Ward identity (AWI), which relates the pseudoscalar density $P^a=\bar \psi \gamma_5 (\tau^a/2)\psi$ to the axial current 
$A_\mu^a=\bar \psi \gamma_\mu\gamma_5 (\tau^a/2)\psi$:
\begin{equation}
\partial_t \sum_{\bf x} \vev{A_0^a({\bf x},t)\cO^a} = 2m_q \sum_{\bf x} \vev{ P^a({\bf x},t)\cO^a}\, ,
\label{eq:AWI}
\end{equation}
where $\cO^a$ is a pseudoscalar Gaussian source and the summation is over the three spatial coordinates.
To obtain the pseudoscalar meson mass ($m_{\rm PS} \equiv E_0$) and the quark mass $m_q$, we fit
the pseudoscalar (PP) and the axial vector (AP) correlators to the functional form
\begin{subequations}
\begin{eqnarray}
\sum_{\bf x} \vev{ P^a({\bf x},t)\cO^a} 
&=& 
\sum_{n=0}^{N-1} C_{n,\rm PP}^2\left[e^{-E_n t} + e^{-E_n (T-t)}\right] \ ,
\label{eq:PP}\\
\sum_{\bf x} \vev{A_0^a({\bf x},t)\cO^a}
&=& 
\sum_{n=0}^{N-1} C_{n,\rm PP} C_{n,\rm AP}\left[e^{-E_n t} - e^{-E_n (T-t)}\right] \ .
\label{eq:AP}
\end{eqnarray}
\label{eq:PPAP}
\end{subequations}
From eqs.~(\ref{eq:AWI}) and~(\ref{eq:PPAP}), we obtain the quark mass
\begin{equation}
m_q = -\frac{E_0}{2}\frac{C_{0,{\rm AP}}}{C_{0,{\rm PP}}}.
\label{eq:mq}
\end{equation}

\begin{figure*}[tb]
\centering
  \includegraphics[width=0.75\textwidth]{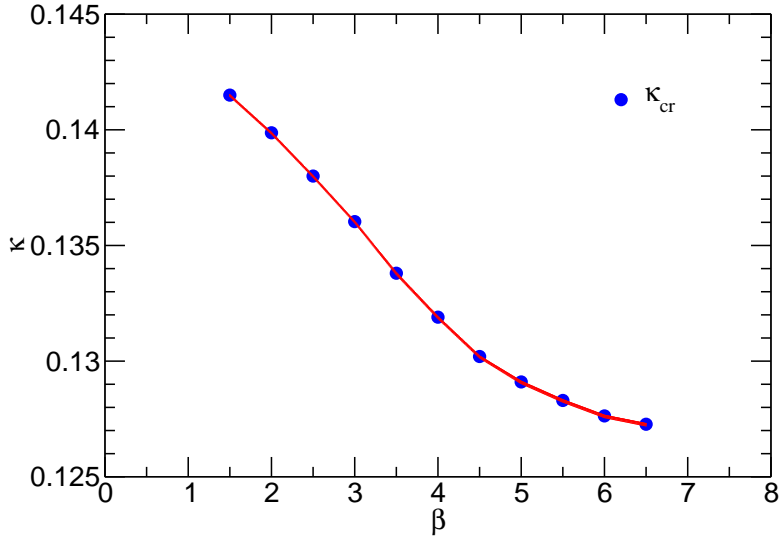}
   \caption{\label{fig:kappac}Critical kappa line determined from $m_q=0$ in \eq{eq:mq}.}
 \end{figure*}

Figure \ref{fig:kappac} shows the $\kappa_\text{cr}(\beta)$ line and Table~\ref{tab:masses} lists the corresponding values, together with the quark masses $m_q$ on $24^4$ volumes.
We found that tuning to the critical line has to be done with increasing precision at strong coupling and on large volumes. Conformal systems in finite volume and at zero fermion mass are volume squeezed, the infrared cut-off is provided by the volume, $\propto 1/L$. Increasing the mass does not have a strong effect as long as the system stays volume squeezed, $m_q \ll 1/L$. However at larger $m_q$ the role of the volume diminishes, the system moves into a mass-deformed regime. The  spectrum in the mass deformed regime  is considerably different than in the volume squeezed one. This phenomena was clearly observed in the finite size scaling analysis of the $N_f=12$ SU(3) system~\cite{Cheng:2013xha}. Chirally broken but volume squeezed systems could show similar behavior~\cite{Ishikawa:2013tua,Ishikawa:2013wf}.

\begin{table}
\centering
\begin{tabular}{lcc ||  lcc}
$\beta$ & $\kappa$ &${\rm m_q}$  &   $\beta$ & $\kappa$ &${\rm m_q}$\\
\hline
1.5 & 0.14150  & 0.0012(3)  & 4.5 & 0.13020  & 0.0020(2)  \\
2.0 & 0.13987  & 0.0007(3)  & 5.0 & 0.12910  &  -0.0007(1)      \\
2.5 & 0.13800  & 0.0018(3)  & 5.5 & 0.12830  & -0.0008(1)  \\
3.0 & 0.13603  & -0.0005(3)   &  6.0 & 0.12763  & 0.0028(1)    \\
3.5 & 0.13380  & 0.0040(3) & 6.5 & 0.12727  & 0.0016(1) \\
4.0 & 0.13190  & 0.0004(2) & & &
\end{tabular}
\caption{Quark mass  values on $24^4$ volumes at $\kappa$ hopping parameter  values used in this work. }
\label{tab:masses}
\end{table}
\begin{figure}[btp]
\center
  \includegraphics[width=0.75\textwidth]{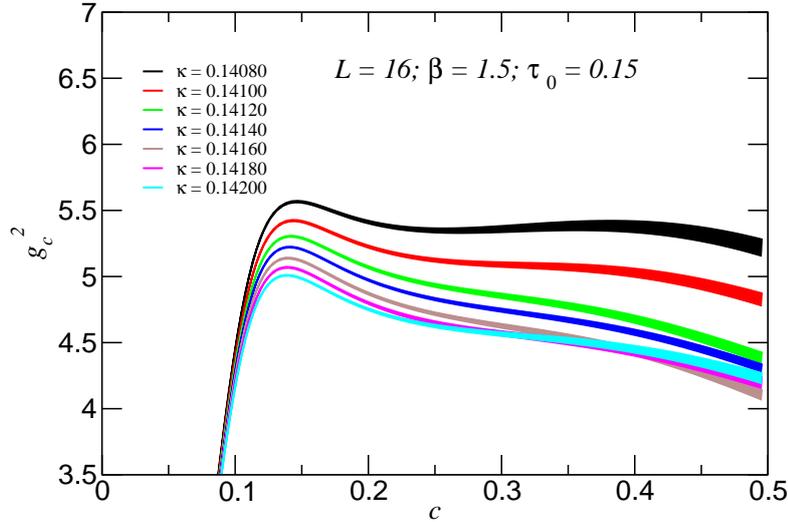}
  \caption{\label{fig:g2_kappa_1616}  The gradient flow coupling \gc as the function of $c=\sqrt{8t}/L$ on $16^4$ volume and $\beta=1.5$ at several, equally spaced,  $\kappa$ values. $\kappa_{\text{cr}}\approx 0.14150$ at this gauge coupling.} \end{figure}
\begin{figure}[btp]
\center
  \includegraphics[width=0.5\textwidth]{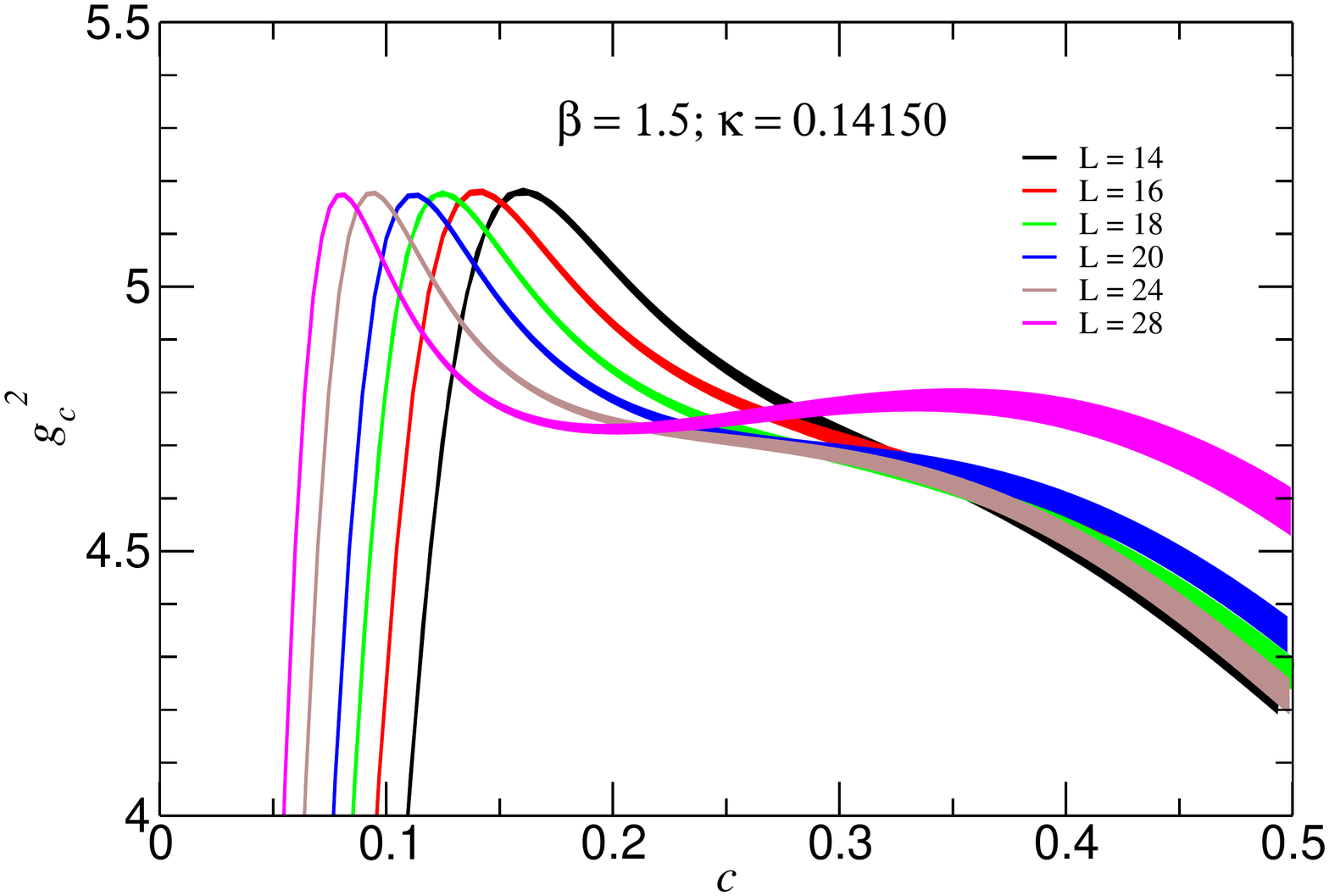}\hfill
  \includegraphics[width=0.5\textwidth]{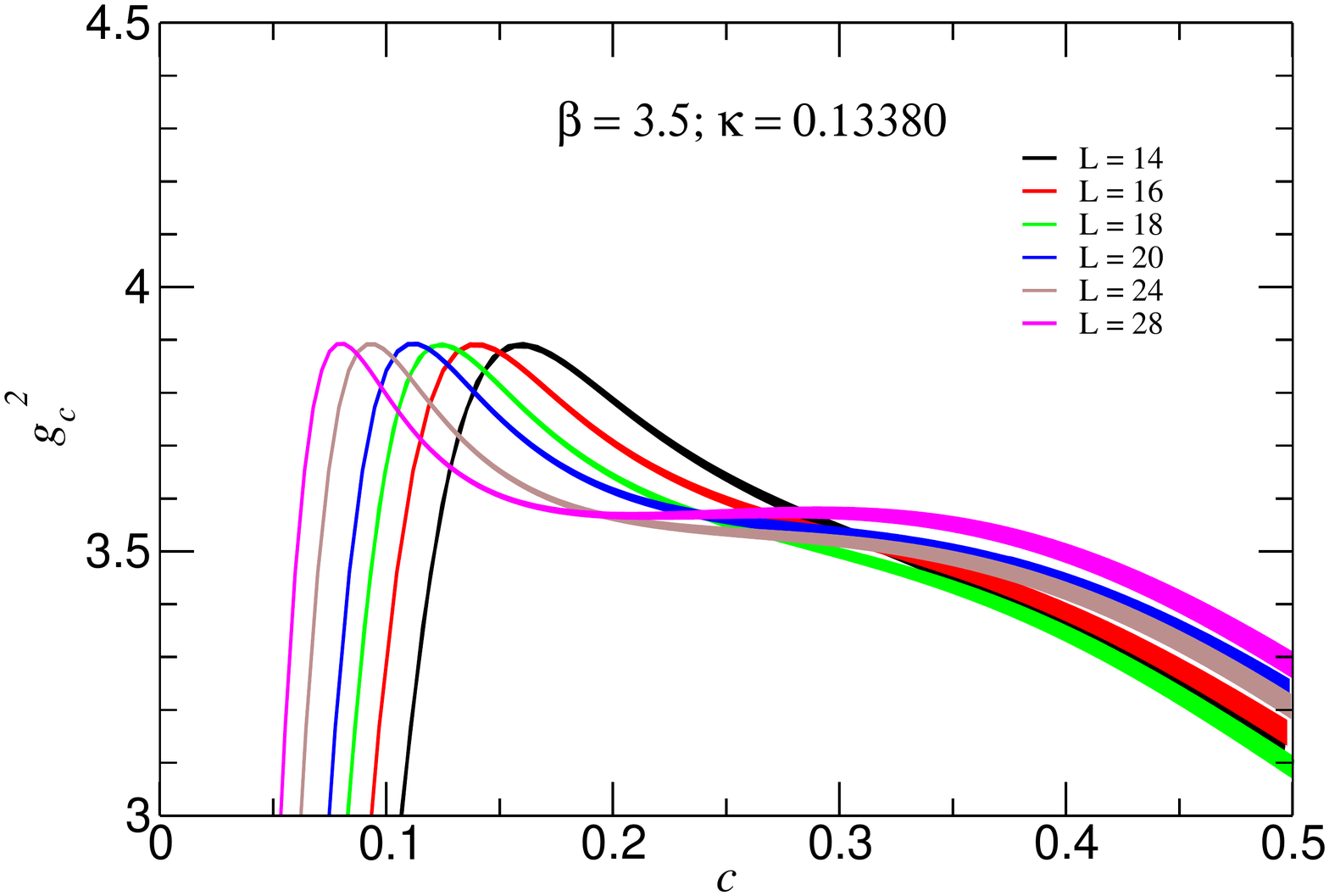}
  \caption{\label{fig:mistuning}  The gradient flow coupling  on   $L/a=14-28$ volumes. The left panel corresponds to $\beta=1.5$, $\kappa=0.14150$, the right to $\beta=3.5$, $\kappa=0.13380$. The strong break between $L/a=24$ and 28 on the left panel indicates that the $28^4$ volume at these parameter values are in the mass deformed regime. }
\end{figure}

In our simulations we observed that incorrect tuning of the fermion mass in either direction can lead to a sudden change in the gradient flow coupling \gGF, leading to severe lattice artifacts. In \fig{fig:g2_kappa_1616} we show the gradient flow coupling  as the function of $c=\sqrt{8t}/L$ at several values of $\kappa$ on $16^4$,  $\beta=1.5$ configurations, both below and above $\kappa_{\text{cr}}\approx 0.14150$. The  $\kappa$ values in  \fig{fig:g2_kappa_1616} are equally spaced. The sudden increase between $\kappa=0.14120$ and 0.14100 signals that the system moved from the volume squeezed to the mass deformed regime. The qualitative change between $\kappa=0.14160$ and 0.14180 indicates the same phenomena, but now in the negative mass region. 

The transition between the volume squeezed and mass deformed regimes depends on the volume. Figure \ref{fig:mistuning} shows the gradient flow coupling  on   $L/a=14-28$ volumes. The left panel corresponds to $\beta=1.5$, $\kappa=0.14150$, the right to $\beta=3.5$, $\kappa=0.13380$. We chose the t-shift $\tau_0$ near its optimal
value, $\tau_0=0.15$ for the left, $\tau_0=0.1$ for the right panel (see sec. \ref{sec:analysis}),  but setting $\tau_0=0.0$ does not change the  qualitative results. While on the right panel \gc evolves smoothly with the volume, on the left panel we observe a strong break between $L/a=24$ and 28. This signals that the $L/a=28$ configurations have moved into the mass-deformed regime and should not be used in the step scaling study. We have seen similar behavior at other $\beta$ values. For example at $\beta=2.0$, $\kappa=0.13984$  we observed that both the $L/a=28$ and 24 volumes are in the mass deformed regime. Even at our final   $\kappa_{\text{cr}}=0.13987$ the $24^4$ configurations could  be mass deformed.   Not surprisingly, smaller volumes can tolerate larger mistuning of $\kappa_{\text{cr}}$, and it is also easier to identify an acceptable $\kappa_{\text{cr}}$ value at weaker gauge  coupling. Overall, while our $\kappa$ values are not perfectly tuned, we can use them on volumes $L/a\le 24$ and $\beta\ge 1.5$. Finding $\kappa_{\text{cr}}$ on larger volumes and stronger couplings would require significant resources and would probably be better done using Schr\"odinger functional boundary conditions. This is beyond our present approach and available resources.

Finally, we mention that we have not found spontaneous chiral symmetry breaking in any of our simulations. Even the $28^4$, $\beta=1.5$, $\kappa=0.14150$ configurations that we discussed above are deconfined; the gradient flow coupling shown in \fig{fig:mistuning}  does not show the linear increase with the gradient flow time as expected in  confining systems.  Approaching and crossing the $\kcr$
line did not cause numerical difficulties even on the largest volume and strongest gauge coupling. We have not observed any first order bulk phase transition at strong coupling, neither an Aoki phase nor  Sharpe-Singleton type first order phase transition~\cite{Aoki:1983qi,Sharpe:1998xm}.

\begin{figure}[btp]
\center
  \includegraphics[width=0.75\textwidth]{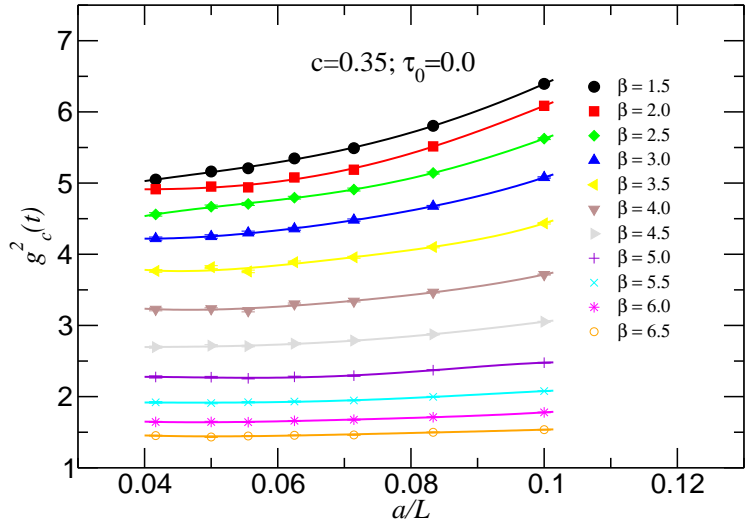}
  \caption{\label{fig:volume_inter}  Volume interpolation of \gc in $a/L$ using fourth order polynomials. }
\end{figure}

\section{\label{sec:analysis}Analysis of the step scaling function} 

For the step scaling function analysis we generated configurations on volumes $8^4$, $10^4$, $12^4$, $14^4$, $16^4$, $18^4$, $20^4$, $24^4$ and $28^4$. We found  that the $8^4$ volumes are too small and introduce large lattice artifacts, while the $28^4$ configurations would require more precise tuning for $\beta\le 2.5$. We do not use those volumes in our final analysis.  We considered 11 gauge coupling values,  spaced 0.5 apart in the $\beta\in[1.5,6.5]$ interval.  We generated $5,000-10,000$ thermalized unit-length molecular dynamics trajectories (MDTU) on the smaller and   $3,000-5,000$   on the larger volumes, evaluating the gradient flow after every 10th MDTU.

In the $L/a\in[10,24]$ range we can form 4 volume-pairs with scale change $s=3/2$: $10 \to 15$, $12 \to 18$, $14 \to 21$, and $16 \to 24$. We do not have $L/a=15$ and 21, instead we interpolate among the 7 volumes $L/a\in[10,24]$. We considered several interpolating functions,  polynomials of degree 3 or 4 in either $\text{log}(L/a)$ or $a/L$. Figure \ref{fig:volume_inter} shows such an interpolation  at $c=0.35$ and  t-shift $\tau_0=0.0$, using fourth order polynomials in $a/L$. Our goal is  to predict a smooth interpolating form which reduces fluctuations among the independent volumes and provide the values of \gc on the missing $L/a=15$ and 21. The predicted values  are  independent of the interpolation details, at least within our statistical errors. We use these interpolated values to determine the step scaling function $\beta_{3/2}(\gc,L)$ of \eq{eq:beta} at each bare coupling $\beta$ and volume pair $L \to 1.5 L$. Since the  $\gc(\beta,L)$ values are independent, we use standard error propagation to estimate errors. 
\begin{figure}[btp]
\center
  \includegraphics[width=0.7\textwidth]{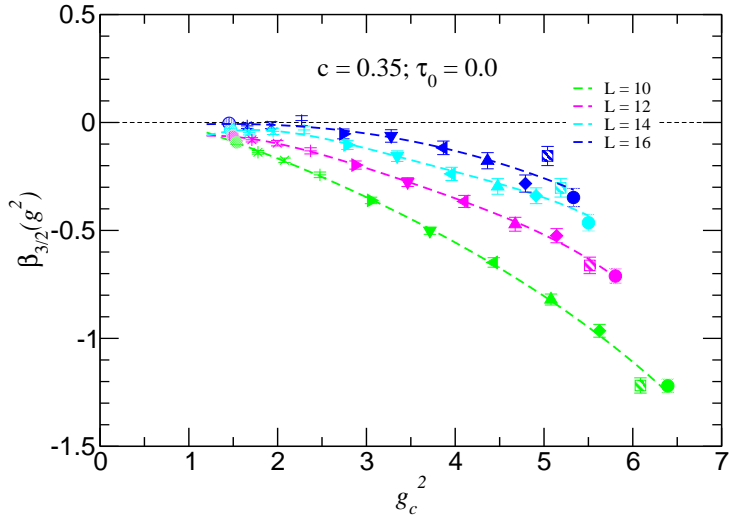}
  \caption{\label{fig:step_0.35_0.0_clov} The   step scaling function at finite $L$,  using volume pairs  $10\to15$, $12\to 18$, $14\to 21$ and $16\to 24$ with   $c=0.35$, $\tau_0=0.0$ and  clover operator discretization.  The different symbols correspond to different $\beta$ values, $\beta=1.5$, 2.0, 2.5, .... 6.5 from right to left. The dashed lines are fourth order interpolating polynomials in $\gc$.}
\end{figure}

Next  we perform  an other interpolation, this time of  $\beta_{3/2}(\gc,L)$ in \gc at fixed $L$. We consider polynomials  of degree 3 or 4  and find that the order of the interpolating polynomial  does not change the predicted $\beta_{3/2}(\gc,L)$ within statistical errors. An example of such an interpolation is shown in \fig{fig:step_0.35_0.0_clov} at $c=0.35$ and $\tau_0=0.0$, using fourth order polynomials for the volume pairs $10 \to 15$, $12 \to 18$, $14 \to 21$, and $16 \to 24$. 
Since in the previous step we interpolated in $L$ at fixed bare coupling and now we interpolate in \gc at fixed L, there is no correlation and standard error analysis applies. 
The different symbols in \fig{fig:step_0.35_0.0_clov} correspond to different $\beta$ values, $\beta=1.5$, 2.0, 2.5, .... 6.5 from right to left. The predictions at different $\beta$ are fully independent, yet they show a fairly smooth dependence on \gc. The only clear exception is the $\beta=2.0$ data set, denoted by dashed squares. On every volume $\beta=2.0$ is off and the deviation increases with the volume. This is a sign of mistuning of the hopping parameter $\kappa$. Any mistuning increases the step scaling function in the continuum limit since the mass is a relevant operator. It appears that the step scaling function is a better indicator of a non-zero bare mass than the fermion mass discussed in \secref{sec:numerical}. Since in our \gc interpolation we include even clearly mistuned $(\beta,\kappa)$ pairs, our resulting step scaling function should be considered as an upper limit of the massless case.

\begin{figure}[btp]
  \includegraphics[width=0.5\textwidth]{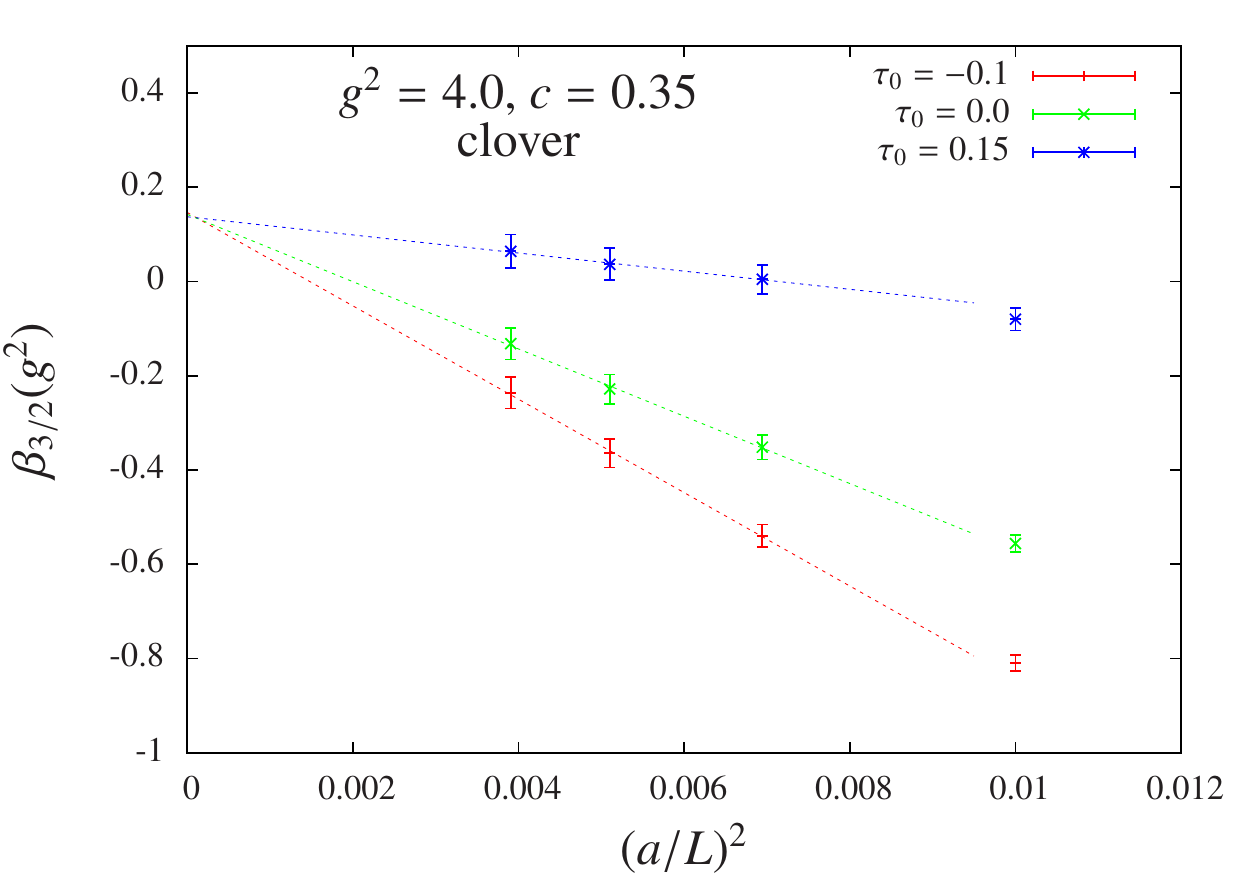}\hfill
  \includegraphics[width=0.5\textwidth]{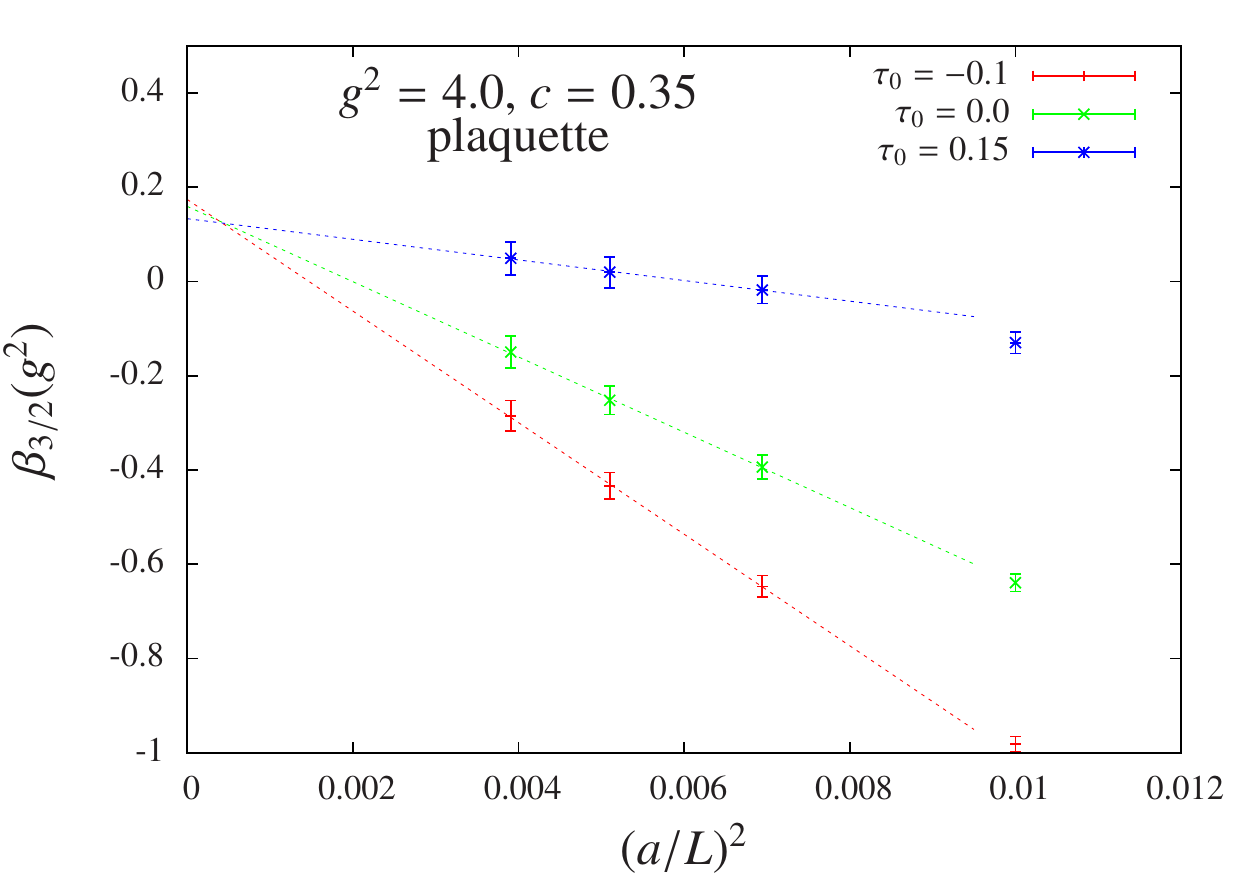}
    \includegraphics[width=0.5\textwidth]{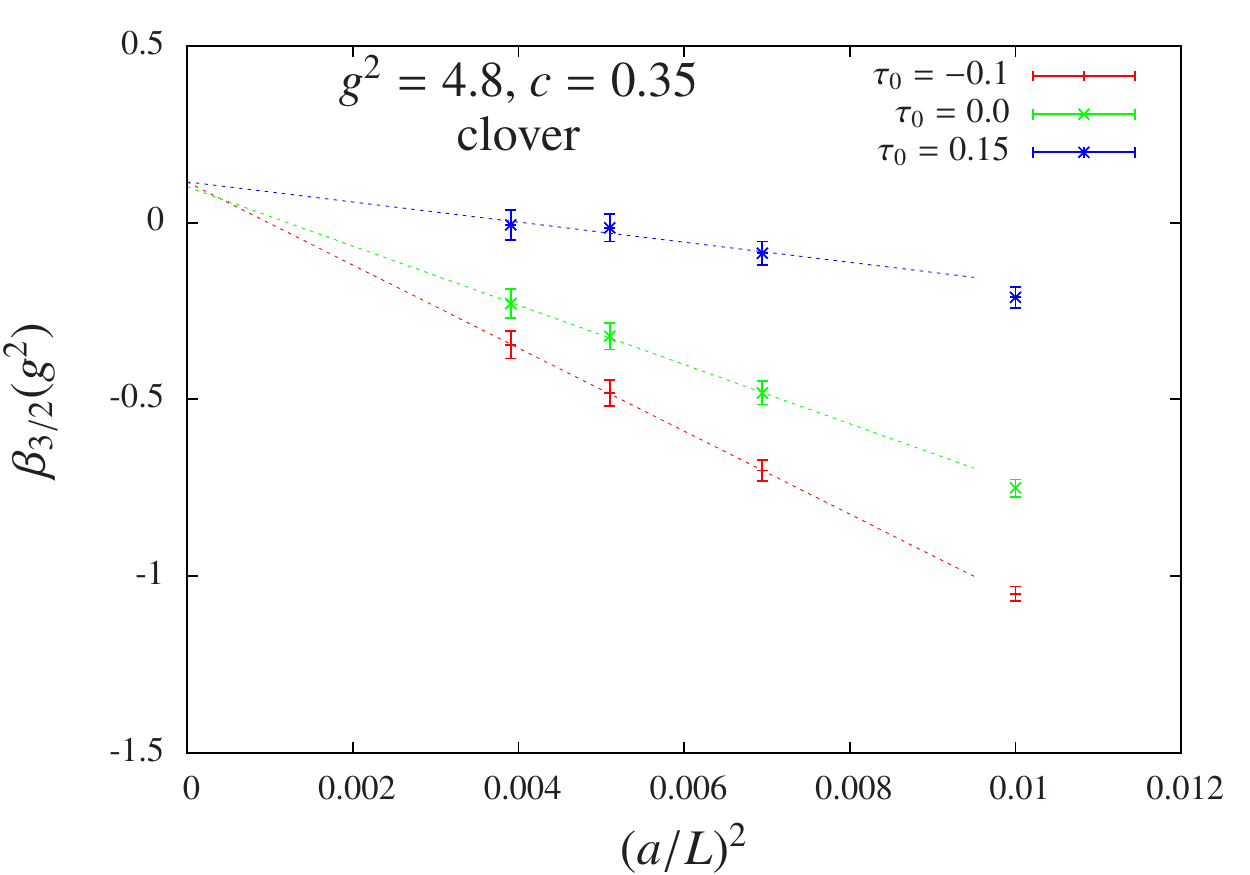}\hfill
  \includegraphics[width=0.5\textwidth]{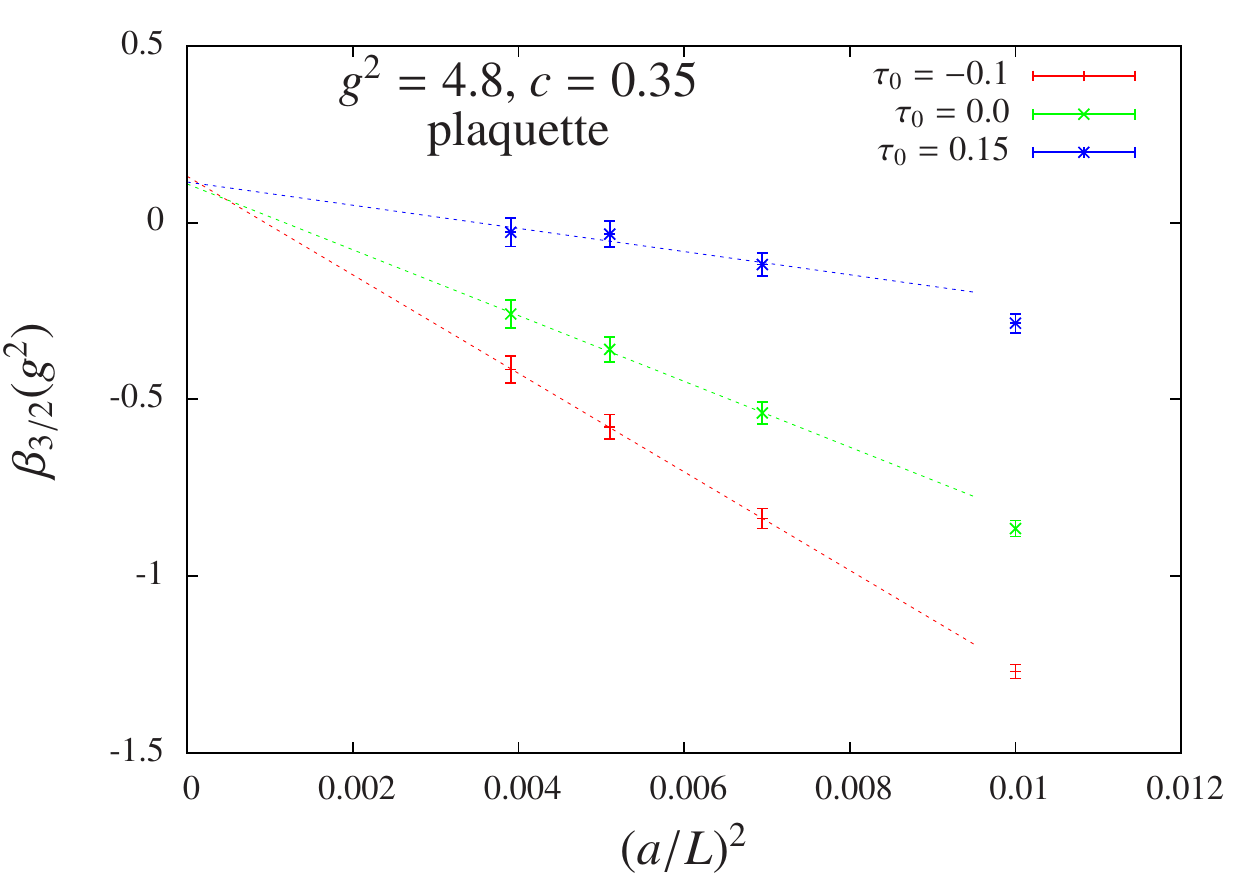}
  \caption{\label{fig:cont_ext_40} Top panels: $(a/L)^2  \to 0$ extrapolation at \gc=4.0, $c=0.35$. The left panel shows data obtained using the clover while the right panel shows data  obtained using the plaquette discretization of the energy density operator. Both panels show results at t-shift $\tau_0 = -0.1$, 0.0 and 0.15. Bottom panels: same as top but for \gc=4.8.}
\end{figure}
\begin{figure}[btp]
\center
  \includegraphics[width=0.7\textwidth]{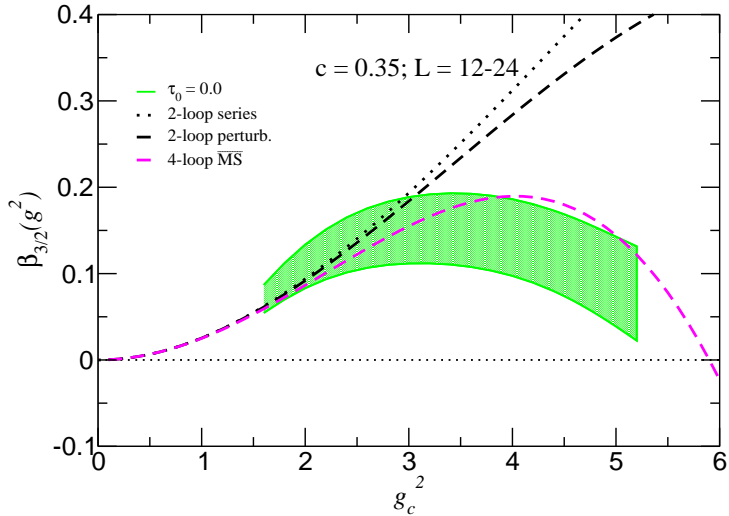}
  \caption{\label{fig:final} The continuum extrapolated $c=0.35$ step scaling function using clover discretization,  $\tau_0=0.0$, and  volume pairs $12\to 18$, $14\to 21$ and $16\to 24$.  The dashed black and magenta lines are the 2-loop and 4-loop \MSbar perturbative predictions obtained by integrating the RG $\beta$ function \eq{eq:2-4-loop}. The dotted line is the  2-loop step scaling function as defined in \refcite{Fodor:2012td}.}
\end{figure}

The last step in determining the continuum limit step scaling function is an extrapolation to the $(a/L) \to 0$  limit. At fixed \gc we use the finite volume step scaling functions determined in the previous interpolating  step.
Figure \ref{fig:cont_ext_40}   shows this continuum extrapolation  at \gc=4.0 and 4.8 with $c=0.35$~\footnote{ We have chosen \gc=4.8 as this is the largest \gc where we can predict $\beta_{3/2}(\gc,L)$ with $\tau_0=0.15$ on all four volume pairs. As we discuss in \secref{sec:cutoff-1} the range of \gc decreases with increasing t-shift.}. Wilson fermions can have $\mathcal{O}(a)$ corrections but nHYP smearing and the clover term with $c_{\text{SW}}=1.0$ significantly reduce those. We found that our values for $\beta_{3/2}(\gc,L)$ are consistent with $(a/L)^2$ dependence. The plots show the continuum extrapolation at several values of the t-shift parameter, $\tau_0=-0.1$, 0.0 and 0.15, as well as the dependence on the energy density operator (clover versus plaquette).  We will compare these in  \secref{sec:cutoff}. 

Using the clover operator,  $\tau_0=0.0$, and  volume pairs $12\to 18$, $14\to 21$ and $16\to 24$ in the range $1.8 \le \gc \le 5.2$ we obtain the continuum extrapolated step scaling function shown in \fig{fig:final}. For reference we  show the universal 2-loop and the \MSbar 4-loop perturbative predictions obtained by integrating the continuum RG $\beta(g^2)$ function~\cite{Ryttov:2010iz}
\begin{equation}
\int_{\gc}^{\gc+\beta_{s}(\gc)} \frac{d\,g^2}{\beta(g^2) } = \text{ln}(s^2) \, , \quad s=3/2.
\label{eq:2-4-loop}
\end{equation}

We also show in \fig{fig:final} the   2-loop value obtained by the series given in \refcite{Fodor:2012td}.  The more than 15\% difference between the two 2-loop discrete $\beta$ functions at \gc = 5.0 show that these perturbative predictions should be used as guidance only.  From now on we will show the perturbative predictions given in \eq{eq:2-4-loop}. We will use the results in \fig{fig:final} as a reference in the comparisons below.

\section{\label{sec:cutoff} Investigating cut-off effects}

In this section we investigate the dependence of the continuum limit  extrapolated step scaling function on the operator used to discretize the energy density,  on the gradient flow defined at different $\tau_0$ t-shift values, and on the volumes used in the  $(a/L) \to 0$ extrapolation. Any dependence on these quantities indicate that the $(a/L) \to 0$ extrapolation did not remove the cut-off corrections and point to higher order $\cO(a^4)$ or non-perturbative effects. Those can be controlled by increasing the   parameter $c = \sqrt{8t} / L$, or by using larger volumes in the continuum extrapolation. 

\subsection{ \label{sec:cutoff-1}Dependence on the t-shift and the gradient flow $c$ parameter} 
\begin{figure}[btp]
  \includegraphics[width=0.5\textwidth]{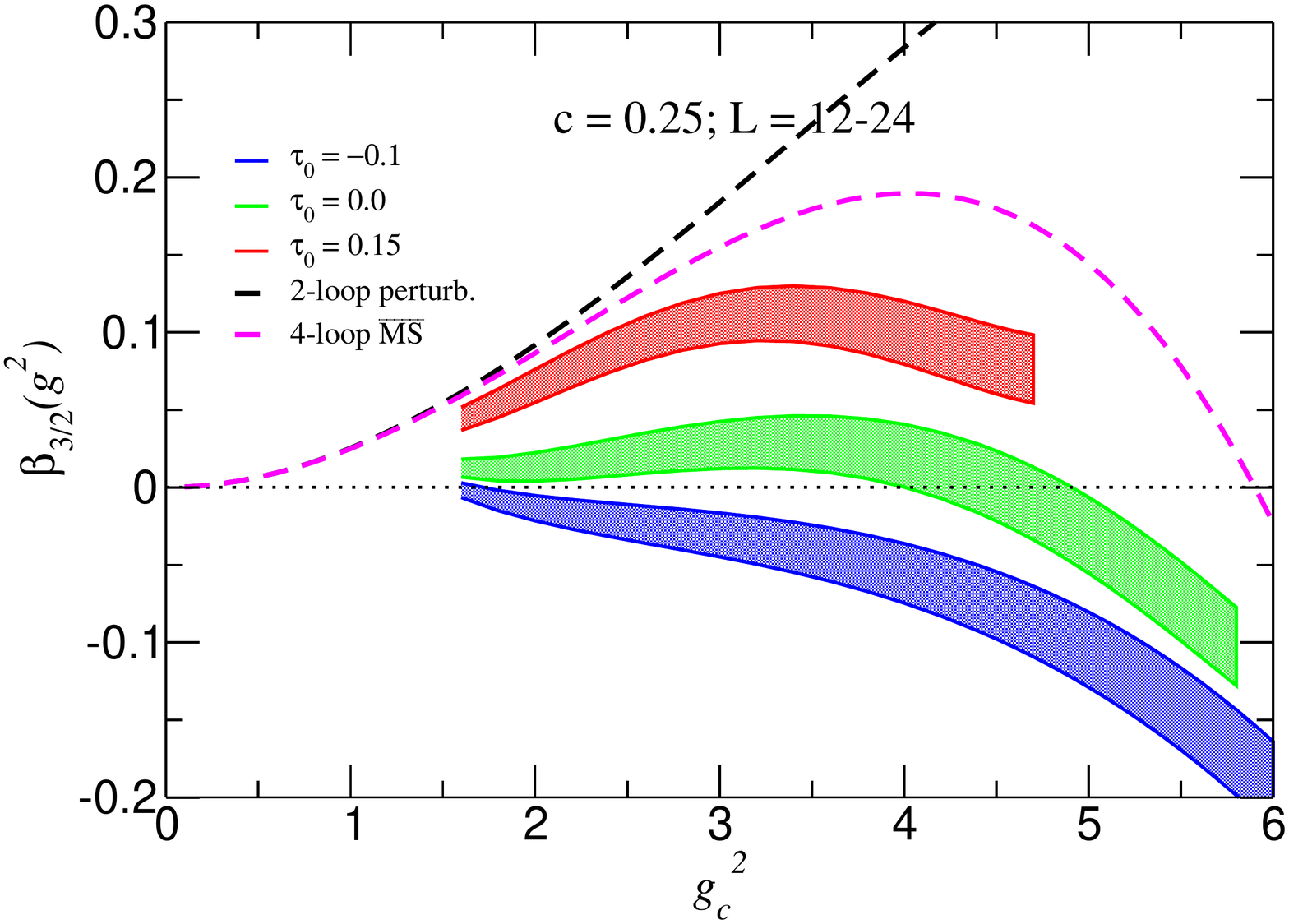}\hfill
  \includegraphics[width=0.5\textwidth]{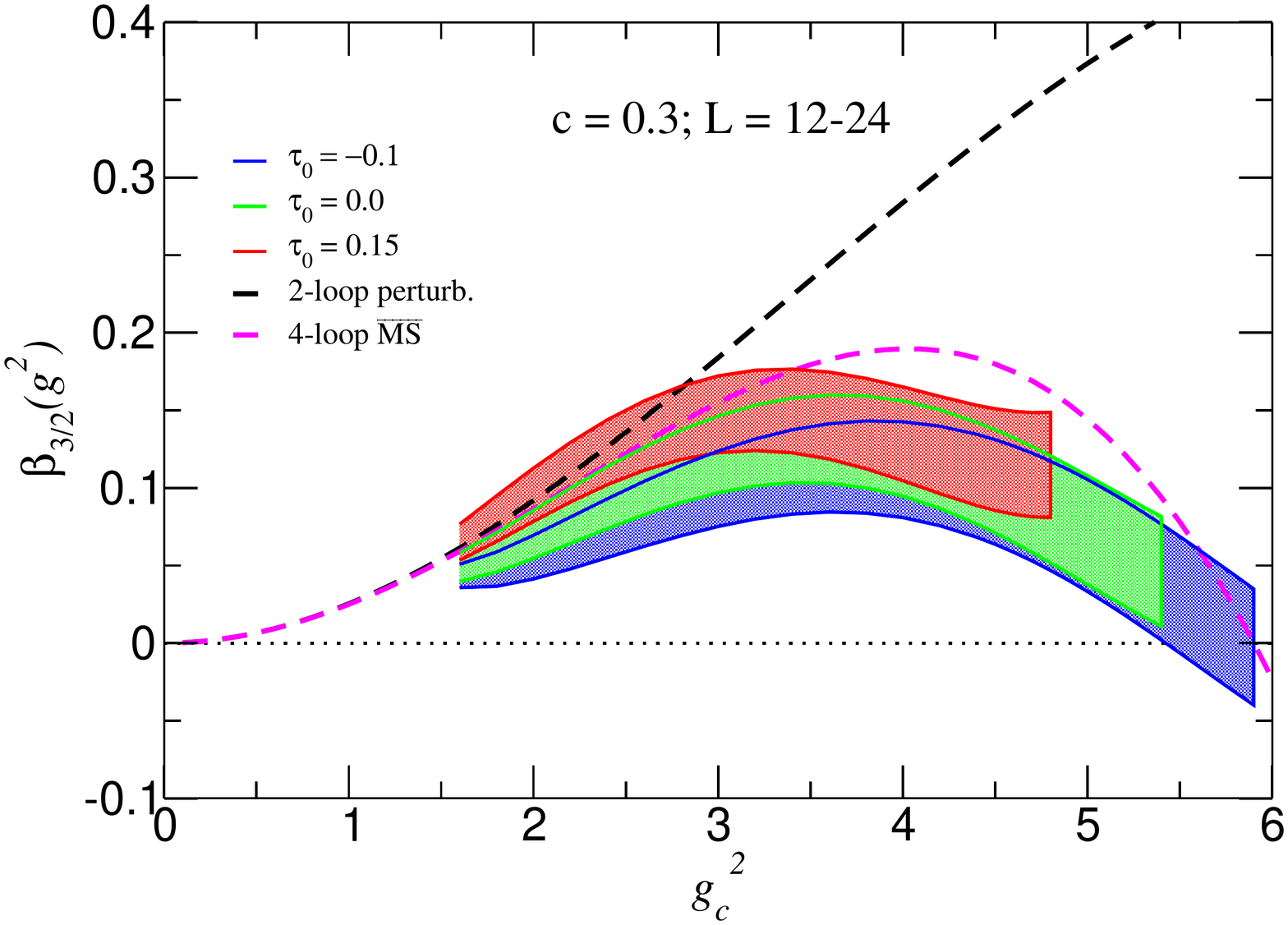}
  \includegraphics[width=0.5\textwidth]{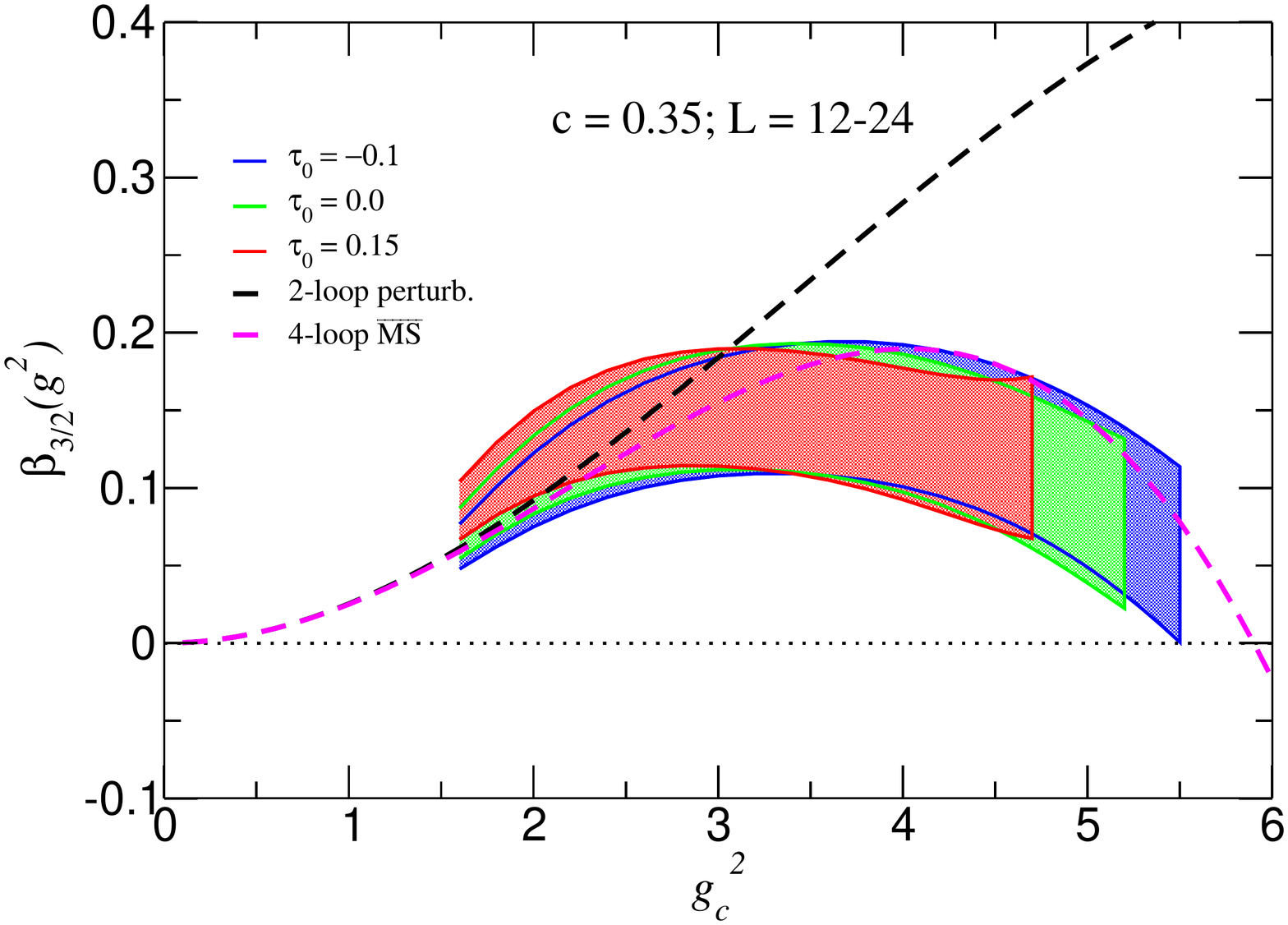}\hfill
  \includegraphics[width=0.5\textwidth]{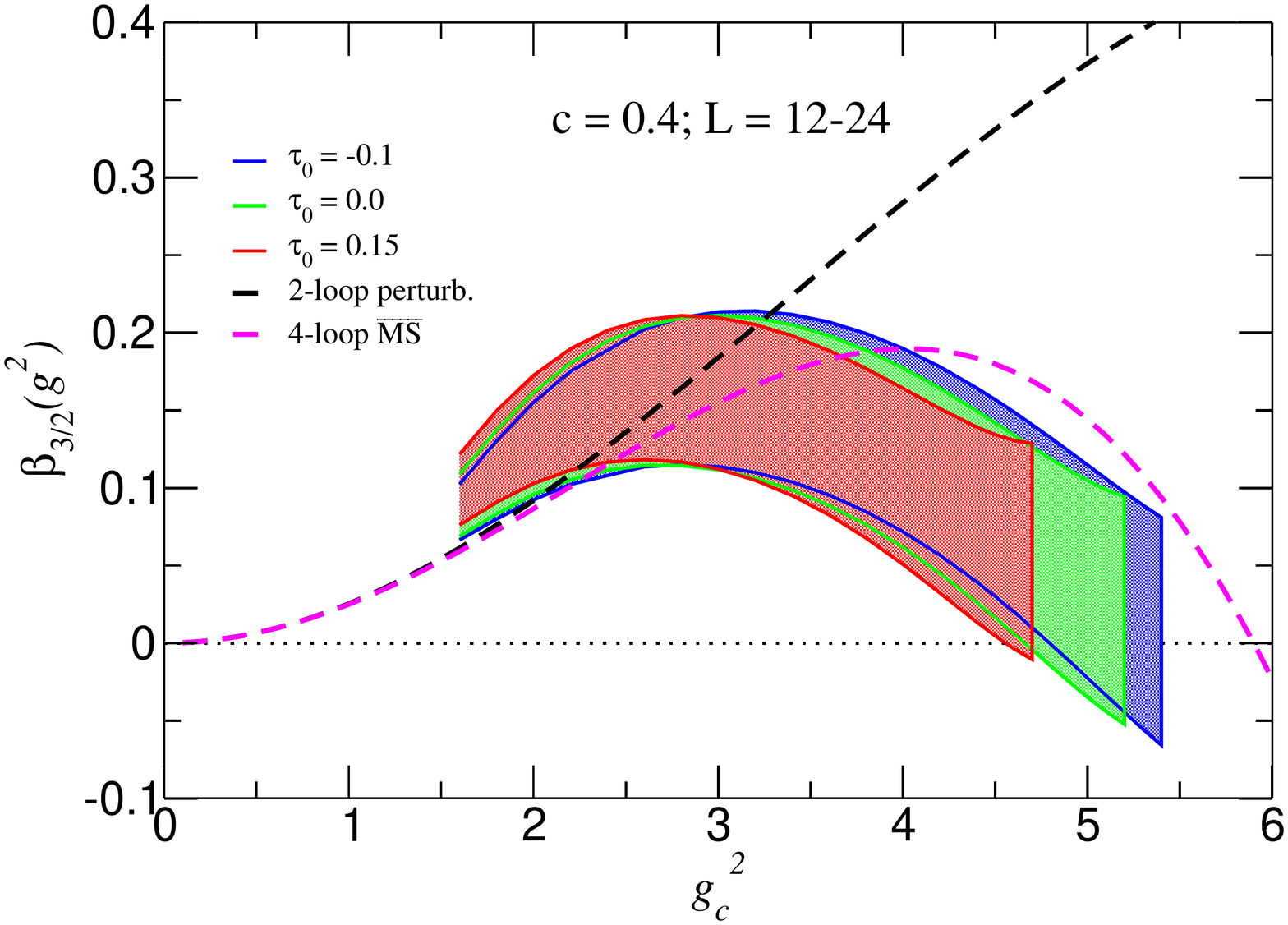}
  \caption{\label{fig:b32_gc_clov_tshift} The continuum limit  extrapolated step scaling function $\beta_{3/2}(\gc)$ with $\tau_0=-0.1$, 0.0 and 0.15 t-shift, using the clover operator. The top row left and right panels correspond to $c=0.25$ and 0.3, the bottom row left and right panels to $c=0.35$ and 0.4.  Significant cut-off effects remain even after the $(a/L)^2\to 0$ extrapolation with $c=0.25$, but they are less noticeable with $c=0.3$ and disappear within statistical errors when $c\ge 0.35$. The dashed black and magenta lines in all panels are the 2-loop and 4-loop \MSbar perturbative predictions.}
\end{figure}

Changing $\tau_0$ in \eq{eq:t-shift}  changes the flow transformation. Deviations in the  continuum extrapolated values at different $\tau_0$ indicate cut-off effects that are not removed by the $(a/L)^2\to 0$ extrapolation.  In \secref{sec:analysis} we have discussed the left panels of  \fig{fig:cont_ext_40}   that  show the continuum extrapolations at \gc=4.0 and 4.8 using the clover operator with $\tau_0=-0.1$, 0.0 and 0.15 and $c=0.35$.  For both \gc   the different t-shifts  predict the   same continuum limit. 
This observation holds at other couplings as well.  The bottom left panel of \fig{fig:b32_gc_clov_tshift} shows the continuum limit extrapolated step scaling functions with $\tau_0=-0.1$, 0.0 and 0.15 and $c=0.35$. The agreement between the three gradient flows  indicates that with $c=0.35$ the $(a/L)^2\to 0$ extrapolation removed most cut-off effects. It is interesting to note that by selecting smaller $\tau_0$ we can reach somewhat larger \gc. It is possible to push this further, selecting $\tau_0=-0.25$ or even $-0.5$, and still find consistent results. However the slope of the $(a/L)^2\to 0$ extrapolation increases with decreasing $\tau_0$ and we do not feel comfortable extending the \gc range this way.

The other panels of \fig{fig:b32_gc_clov_tshift} show the $(a/L)^2\to 0$ extrapolated step scaling functions with  $c=0.25$, 0.3 and 0.4 with the same $\tau_0=-0.1$, 0.0 and 0.15 values. The statistical errors are significantly smaller at $c=0.25$ on the top left panel.  While the dependence of $\beta_{3/2}(\gc,L)$ on $(a/L)^2$ is still reasonably  linear,   the $(a/L)^2\to 0$ extrapolated values depend very strongly on the t-shift parameter.   Clearly the $(a/L)^2\to 0$ extrapolation does not remove the cut-off effects at $c=0.25$. Had we chosen to work with $\tau_0=0.0$ only, we would not have observed this inconsistency and could have predicted an incorrect step scaling function. 

The step scaling function at $c=0.3$,  on the top right panel of \fig{fig:b32_gc_clov_tshift}, shows much better consistency between different t-shift values than $c=0.25$. For $c= 0.35$ and 0.4, shown on the bottom two panels, we do not observe any cut-off dependence within our statistical uncertainty. 
Different $c$ values define different RG schemes and their corresponding step scaling functions do not have to agree. Nevertheless we see in \fig{fig:b32_gc_clov_tshift}  that the continuum limit extrapolated step scaling functions with $c\ge 0.3$ are very close.

All panels of \fig{fig:b32_gc_clov_tshift} used the clover operator. The plaquette operator gives very similar results.

\subsection{ Dependence on the energy density operator} 
\begin{figure}[btp]
  \includegraphics[width=0.5\textwidth]{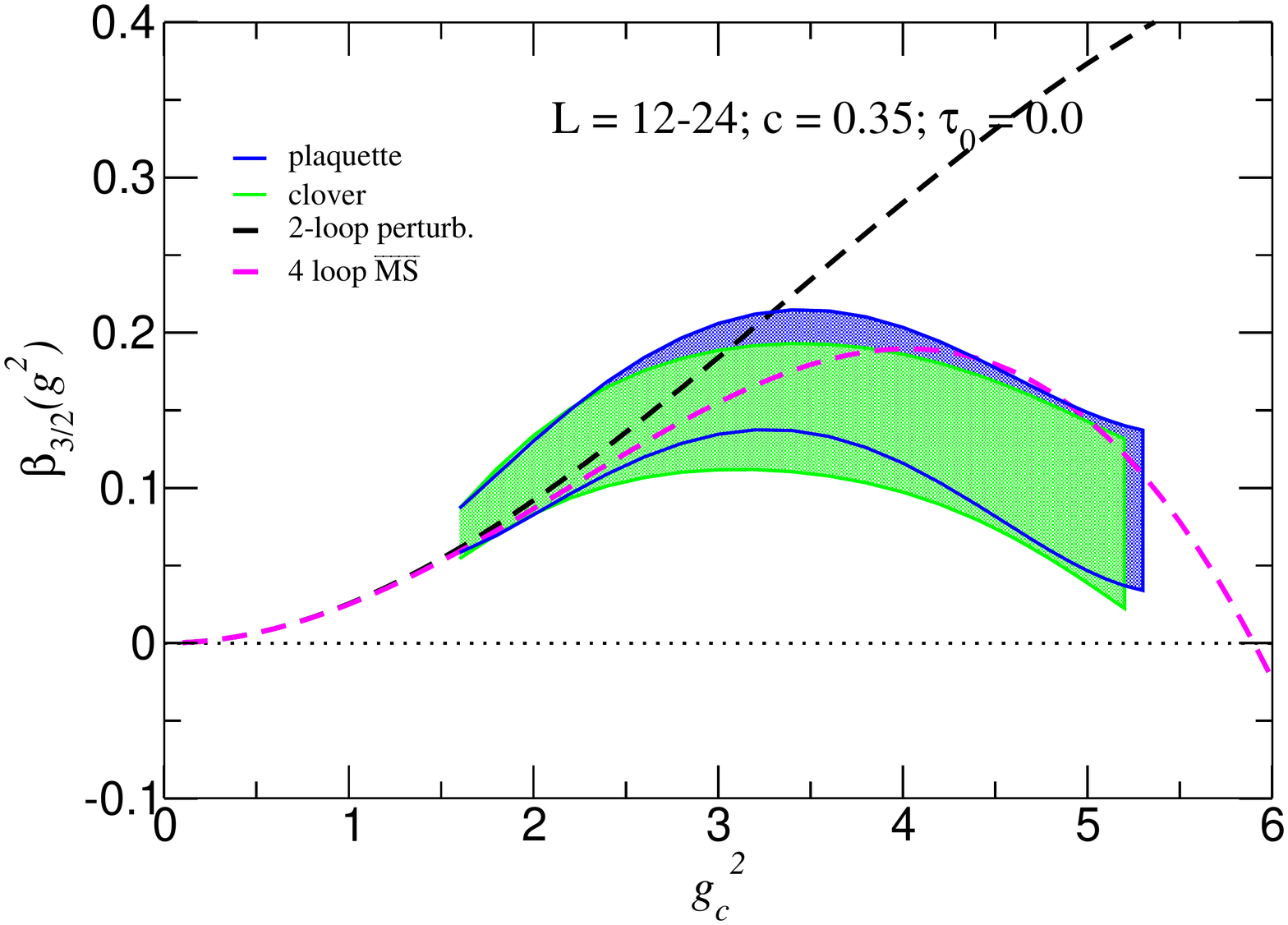}\hfill
  \includegraphics[width=0.5\textwidth]{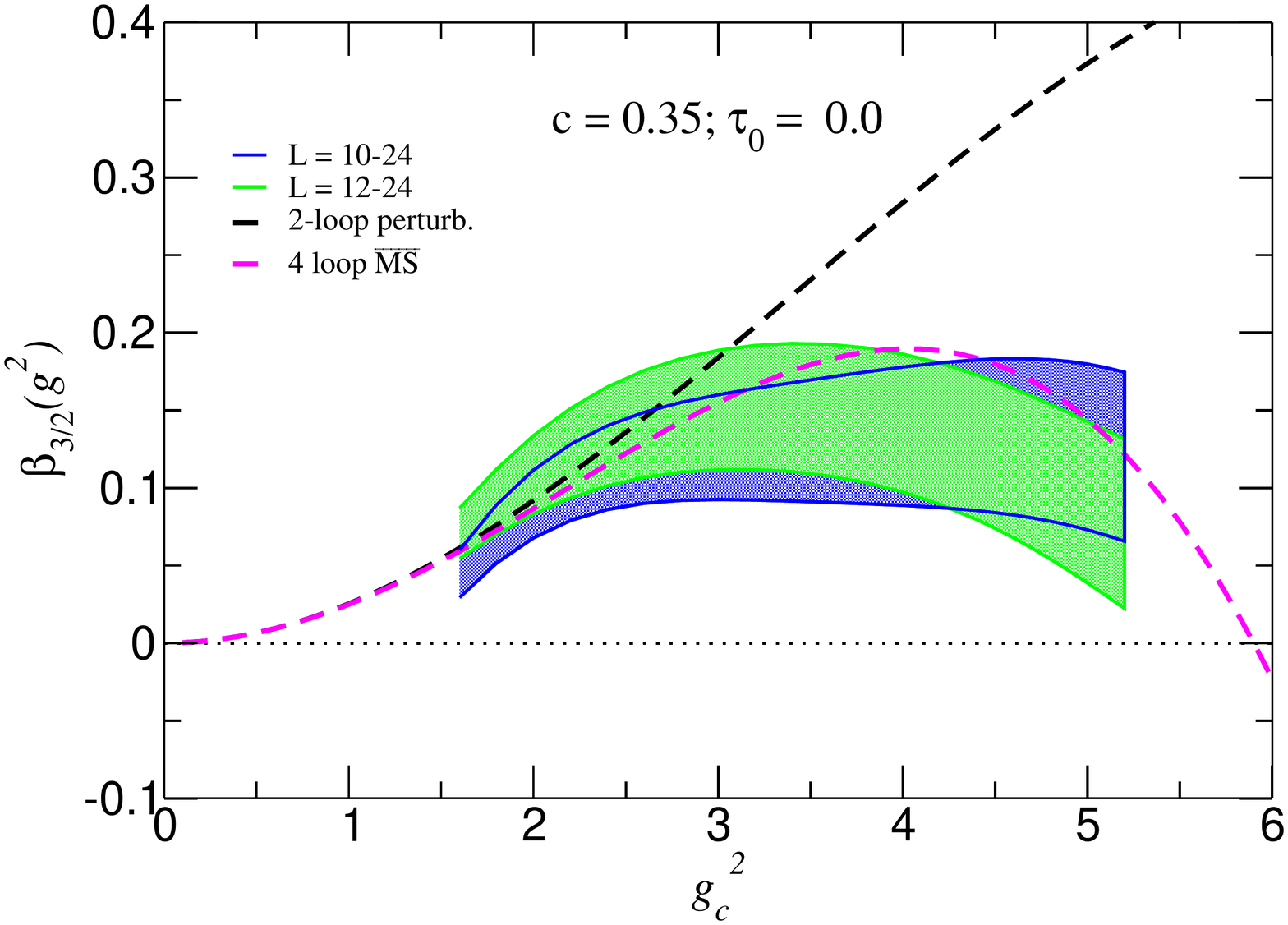}
  \caption{\label{fig:b32_gc_clov_plaq} The $(a/L)\to 0$ extrapolated step scaling function $\beta_{3/2}(\gc)$ with $\tau_0=0.0$, $c=0.35$. The left panel compares the clover and plaquette discretization, while the right panel compares  $(a/L)^2 \to 0$  extrapolations using all 4 or only the 3 largest volume pairs. The dashed black and magenta lines are the 2-loop and 4-loop \MSbar perturbative predictions.}
\end{figure}

The energy density $E(t) = -\frac{1}{2}\mbox{ReTr}\left[G_{\mu\nu}(t) G^{\mu\nu}(t)\right]$ in \eq{eq:gGF} can be discretized in many different ways. The two most frequently used discretizations are based on the plaquette and  clover operators~\cite{Luscher:2010iy}. The different  discretizations have different cut-off effects and comparing the two is an other way to verify that the $(a/L)^2 \to 0$  extrapolation  predicts the continuum step scaling function~\cite{Lin:2014fxa}.

The right panels of  \fig{fig:cont_ext_40}  show the continuum extrapolations at \gc=4.0 and 4.8 using the plaquette operator and the corresponding $C(L,c)$ values in \eq{eq:pert_g2}. Comparing the right and left panels we see that while the cut-off effects at given t-shift are different, the continuum extrapolated values are identical with the two operators. This observation holds at other \gc values as well.  On the left panel of \fig{fig:b32_gc_clov_plaq}  we overlay the continuum extrapolated step scaling functions predicted by the plaquette and clover operators with $c=0.35$, $\tau_0=0.0$. The agreement between the  two operators  indicates that with $c=0.35$ the $(a/L)^2\to 0$ extrapolation  removes most cut-off effects both with clover and plaquette discretization.

Repeating the same comparison at $c=0.25$ shows deviation between the two operators, but the effect is much less striking than comparing different t-shifts as in \fig{fig:b32_gc_clov_tshift}.

\subsection{ Dependence on  the lattice volume }

Our volumes, while not small, are smaller than \refcite{Fodor:2015zna}. It is important to verify that they are large enough to predict the correct continuum limit. The various consistency checks discussed above show that with $c\ge 0.35$ this is the case.

In every analysis until now we extrapolated to the continuum limit using volumes $L/a\ge 12$. As one can infer from \fig{fig:cont_ext_40}  the $L/a=10 \to 15$ prediction for $\beta_{3/2}(\gc)$ is fairly consistent with that. (The $L/a=8\to12$ predictions, on the other hand, are significantly off. We do not show those in \fig{fig:cont_ext_40}   as we do not use the $8^4$ volumes anywhere in the analysis.)
Including the volume pair $L/a=10 \to 15$ in the  $(a/L)^2\to 0$ extrapolation indeed has negligible effect on the step scaling function as the right panel of \fig{fig:b32_gc_clov_plaq} shows.  The agreement is well within statistical errors.

Looking at \fig{fig:step_0.35_0.0_clov} it is clear that \gc depends only weakly on $L$ at fixed bare coupling $\beta$. This is due to the very slow running of the coupling and is independent of the action or RG scheme used.  The $(a/L) \to 0$  limit at larger \gc values on our volumes covers a relatively small region in  bare couplings, though the interpolation of $\beta_{3/2}(\gc,L)$  in \gc connects the weak and strong coupling regions. In any case we should consider the  question if the usual  $(a/L)^2 \to 0$  extrapolation is acceptable in a slowly running, near-conformal system, or it would make more sense to  allow an arbitrary power-like dependence in $L$, as was advocated in refs. \cite{DeGrand:2012yq,DeGrand:2012qa}. While at or near an IRFP one might indeed find a different $L$ dependence, we do not see any indication of that in our study. Different t-shift values and operators predict the same continuum step scaling function, supporting the validity of the $(a/L)^2 \to 0$ extrapolation. Even in the study of the 12-flavor  system an  $(a/L)^2 \to 0$ extrapolation  of the IRFP is satisfactory~\cite{Cheng:2014jba}. Nevertheless there is a way to mitigate this issue. As \fig{fig:cont_ext_40} shows it is possible to change the  $(a/L)$ dependence of $\beta_{3/2}(\gc)$ by changing the  $\tau_0$ t-shift value. With the clover operator at $c=0.35$ there is practically no dependence on $L$ with $\tau_0=0.2$ in the $\gc \ge 3.5$ region, so the actual functional form of the $L$ dependence becomes irrelevant. With the plaquette operator this optimal  value is also close to $\tau_0 = 0.2$. In both cases we find  the same continuum extrapolated step scaling function as with other $\tau_0$ values.

\section{\label{sec:conclusion} Conclusion and discussion}

We investigated  the step scaling (discrete RG $\beta$)  function of the 2-flavor SU(3) sextet model,  one of the leading candidates of BSM models that describe the Higgs  boson as a composite particle. Using nHYP smeared clover-Wilson fermions we were able to cover the range $\gc \lesssim 5.5$ where we find that the step scaling function follows closely the 4-loop (non-universal) \MSbar prediction that predicts an IRFP at  $g^2 \approx 5.90$.   Lattice calculations can suffer from cut-off effects. We made special efforts to investigate them by using different discretizations and gradient flow transformations. We found that most cut-off effects are removed if the gradient flow parameter $c\ge0.35$, leading to consistent results. 

An independent study of the step scaling   function of this model  was reported recently~\cite{Fodor:2015zna}.  The results,  using staggered fermions  and the same  gradient flow method,  agree with perturbation theory at small couplings and show  a downward deviation relative to the 2-loop $\beta$-function at stronger couplings. However this downward deviation is much weaker (in fact consistent with the 2-loop prediction of \eq{eq:2-4-loop}) than what we observe with Wilson fermions.  Since we use the same renormalized coupling to calculate the step scaling function as \refcite{Fodor:2015zna}, predictions from different lattice actions are expected to agree. However, the results show an over $3 \sigma$ tension at strong couplings between the two works.

At present we do not understand the source of this disagreement between the staggered and Wilson fermion results. This problem should be investigated in the future on larger volumes, perhaps using different lattice formulations or even different lattice methods.

In closing we would like to pose a question regarding universality that might shed light to this puzzle.   The concept of universality is the cornerstone of critical phenomena and lattice QCD calculations, yet it has not been investigated in connection with 4 dimensional many-fermion conformal systems.
 Universality requires that the lattice actions are local,  have the same field content, and the same symmetries. If two lattice actions satisfy these conditions and are in the vicinity (basin of attraction)  of the same fixed point, their critical properties (continuum limit) are identical.   The two widely used lattice fermion formulations, Wilson and staggered fermions, have very different symmetries. The former completely breaks chiral symmetry while the latter has a remnant U(1) chiral symmetry. Nevertheless in QCD-like systems, around the perturbative Gaussian fixed point, one can prove that they differ only in irrelevant operators and describe the same continuum quantum field theory. In conformal systems the fate of the two fermion formulations is not obvious  in the basin of attraction of the infrared fixed point,  as irrelevant operators can become relevant and vice versa at different fixed points. 

Similar questions have been studied extensively in 3 dimensional  spin systems. A 5-loop $\epsilon$-expansion of the $O(n_1) \oplus O(n_2)$ model shows that for $N=n_1+n_2\ge 3$ the $O(N)$ fixed point is unstable, the critical properties of the system are controlled either by a biconal or a tetra-critical decoupled fixed point~\cite{Calabrese:2002bm}. 
3 dimensional spin models are very different from 4 dimensional gauge-fermion systems, but the richness of the phase structure of the 3 dimensional spin systems suggests that in 4 dimension we might find unexpected results as well. 
If it turned out that Wilson and staggered fermions have different IRFPs in a conformal system, it would not be surprising to find different step scaling function close to the IRFP. 

We cannot conclude that the 2-flavor sextet system is conformal, nor that it is  chirally broken.   However our step scaling function approaches zero very closely. If the system is not conformal it has to be slowly walking and its  IR dynamics could be influenced by a nearby IRFP.  The question of universality between staggered and Wilson fermions arises again. 

Considering the possible phenomenological importance of this system, it is imperative to understand and resolve these issues.

\section*{Acknowledgments} 
We are indebted to B.~Svetitsky for his participation in the early stages of this project and his many insightful comments on the final manuscript. We thank Y.~Shamir for his valuable suggestions about using the NDS action and for many helpful discussions. We thank A.~Ramos and C.~J.~D.~Lin for their suggestion to compare the plaquette and clover discretized operators. We used the NDS Wilson fermion code developed by Y.~Shamir,  based in part on the MILC Collaboration's public lattice gauge theory software\footnote{\texttt{http://www.physics.utah.edu/$\sim$detar/milc/}}.
 Numerical calculations were carried out on the HEP-TH and Janus clusters, partially funded by NSF Grant No.~CNS-0821794  at the University of Colorado; at Fermilab under the auspices of USQCD supported by the DOE; and at the Gordon  Computing Center through XSEDE supported by National Science Foundation Grant No.~OCI-1053575.
 This research was partially supported by the U.S.~Department of Energy (DOE) through Grant No.~DE-SC0010005 (A.~H., Y.~L.) .

\bibliographystyle{utphys}
\bibliography{sextet}

\providecommand{\href}[2]{#2}\begingroup\raggedright\begin{thebibliography}{10}

\bibitem{Aad:2015yza}
ATLAS Collaboration, G.~Aad {\em et al.}, ``{Search for a new resonance
  decaying to a $W$ or $Z$ boson and a Higgs boson in the $\ell \ell/ \ell \nu/
  \nu \nu + b \bar{b}$ final states with the ATLAS Detector}'',
  \href{http://dx.doi.org/10.1140/epjc/s10052-015-3474-x}{{\em Eur. Phys. J.}
  {\bf C75} (2015) no.~6, 263},
\href{http://arxiv.org/abs/1503.08089}{{\tt arXiv:1503.08089 [hep-ex]}}.

\bibitem{Aad:2015owa}
ATLAS Collaboration, G.~Aad {\em et al.}, ``{Search for high-mass diboson
  resonances with boson-tagged jets in proton-proton collisions at $\sqrt{s}$ =
  8 TeV with the ATLAS detector}'',
\href{http://arxiv.org/abs/1506.00962}{{\tt arXiv:1506.00962 [hep-ex]}}.

\bibitem{Hong:2004td}
D.~K. Hong, S.~D.~H. Hsu, and F.~Sannino, ``{Composite Higgs from higher
  representations}'',
  \href{http://dx.doi.org/10.1016/j.physletb.2004.07.007}{{\em Phys. Lett.}
  {\bf B597} (2004)  89--93},
\href{http://arxiv.org/abs/hep-ph/0406200}{{\tt arXiv:hep-ph/0406200
  [hep-ph]}}.

\bibitem{Dietrich:2005jn}
D.~D. Dietrich, F.~Sannino, and K.~Tuominen, ``{Light composite Higgs from
  higher representations versus electroweak precision measurements: Predictions
  for CERN LHC}'', \href{http://dx.doi.org/10.1103/PhysRevD.72.055001}{{\em
  Phys. Rev.} {\bf D72} (2005)  055001},
\href{http://arxiv.org/abs/hep-ph/0505059}{{\tt arXiv:hep-ph/0505059
  [hep-ph]}}.

\bibitem{Dietrich:2006cm}
D.~D. Dietrich and F.~Sannino, ``{Conformal window of SU(N) gauge theories with
  fermions in higher dimensional representations}'',
  \href{http://dx.doi.org/10.1103/PhysRevD.75.085018}{{\em Phys. Rev.} {\bf
  D75} (2007)  085018},
\href{http://arxiv.org/abs/hep-ph/0611341}{{\tt arXiv:hep-ph/0611341
  [hep-ph]}}.

\bibitem{Arbey:2015exa}
A.~Arbey, G.~Cacciapaglia, H.~Cai, A.~Deandrea, S.~Le~Corre, and F.~Sannino,
  ``{Fundamental Composite Electroweak Dynamics: Status at the LHC}'',
\href{http://arxiv.org/abs/1502.04718}{{\tt arXiv:1502.04718 [hep-ph]}}.

\bibitem{Fodor:2012ni}
Z.~Fodor, K.~Holland, J.~Kuti, D.~Nogradi, C.~Schroeder, {\em et al.}, ``{The
  sextet gauge model, light Higgs, and the dilaton}'', {\em PoS} {\bf
  LATTICE2012} (2012)  024,
\href{http://arxiv.org/abs/1211.6164}{{\tt arXiv:1211.6164 [hep-lat]}}.

\bibitem{Fodor:2012ty}
Z.~Fodor, K.~Holland, J.~Kuti, D.~Nogradi, C.~Schroeder, and C.~H. Wong, ``{Can
  the nearly conformal sextet gauge model hide the Higgs impostor?}'',
  \href{http://dx.doi.org/10.1016/j.physletb.2012.10.079}{{\em Phys. Lett.}
  {\bf B718} (2012)  657--666},
\href{http://arxiv.org/abs/1209.0391}{{\tt arXiv:1209.0391 [hep-lat]}}.

\bibitem{Fodor:2014pqa}
Z.~Fodor, K.~Holland, J.~Kuti, D.~Nogradi, and C.~H. Wong, ``{Can a light Higgs
  impostor hide in composite gauge models?}'',
  \href{http://pos.sissa.it/archive/conferences/187/062/LATTICE
  2013_062.pdf}{{\em PoS} {\bf LATTICE 2013} (2014)  062},
  \href{http://arxiv.org/abs/1401.2176}{{\tt arXiv:1401.2176}}.

\bibitem{Foadi:2012bb}
R.~Foadi, M.~T. Frandsen, and F.~Sannino, ``{125 GeV Higgs boson from a not so
  light technicolor scalar}'',
  \href{http://dx.doi.org/10.1103/PhysRevD.87.095001}{{\em Phys. Rev.} {\bf
  D87} (2013) no.~9, 095001},
\href{http://arxiv.org/abs/1211.1083}{{\tt arXiv:1211.1083 [hep-ph]}}.

\bibitem{Fodor:2015vwa}
Z.~Fodor, K.~Holland, J.~Kuti, S.~Mondal, D.~Nogradi, and C.~H. Wong, ``{Toward
  the minimal realization of a light composite Higgs}'', {\em PoS} {\bf
  LATTICE2014} (2015)  244,
\href{http://arxiv.org/abs/1502.00028}{{\tt arXiv:1502.00028 [hep-lat]}}.

\bibitem{Fukano:2010yv}
H.~S. Fukano and F.~Sannino, ``{Conformal Window of Gauge Theories with
  Four-Fermion Interactions and Ideal Walking}'',
  \href{http://dx.doi.org/10.1103/PhysRevD.82.035021}{{\em Phys. Rev.} {\bf
  D82} (2010)  035021},
\href{http://arxiv.org/abs/1005.3340}{{\tt arXiv:1005.3340 [hep-ph]}}.

\bibitem{Ryttov:2010iz}
T.~A. Ryttov and R.~Shrock, ``Higher-loop corrections to the infrared evolution
  of a gauge theory with fermions'',
  \href{http://dx.doi.org/10.1103/PhysRevD.83.056011}{{\em Phys. Rev.} {\bf
  D83} (2011)  056011}, \href{http://arxiv.org/abs/1011.4542}{{\tt
  arXiv:1011.4542}}.

\bibitem{Shamir:2008pb}
Y.~Shamir, B.~Svetitsky, and T.~DeGrand, ``{Zero of the discrete beta function
  in SU(3) lattice gauge theory with color sextet fermions}'',
  \href{http://dx.doi.org/10.1103/PhysRevD.78.031502}{{\em Phys.Rev.} {\bf D78}
  (2008)  031502},
\href{http://arxiv.org/abs/0803.1707}{{\tt arXiv:0803.1707 [hep-lat]}}.

\bibitem{DeGrand:2010na}
T.~DeGrand, Y.~Shamir, and B.~Svetitsky, ``{Running coupling and mass anomalous
  dimension of SU(3) gauge theory with two flavors of symmetric-representation
  fermions}'', \href{http://dx.doi.org/10.1103/PhysRevD.82.054503}{{\em
  Phys.Rev.} {\bf D82} (2010)  054503},
\href{http://arxiv.org/abs/1006.0707}{{\tt arXiv:1006.0707 [hep-lat]}}.

\bibitem{DeGrand:2012yq}
T.~DeGrand, Y.~Shamir, and B.~Svetitsky, ``{Mass anomalous dimension in sextet
  QCD}'', \href{http://dx.doi.org/10.1103/PhysRevD.87.074507}{{\em Phys.Rev.}
  {\bf D87} (2013)  074507},
\href{http://arxiv.org/abs/1201.0935}{{\tt arXiv:1201.0935 [hep-lat]}}.

\bibitem{Sinclair:2014cga}
D.~K. Sinclair and J.~B. Kogut, ``{Models of Walking Technicolor on the
  Lattice}'', {\em PoS} {\bf LATTICE2014} (2014)  239,
\href{http://arxiv.org/abs/1410.8494}{{\tt arXiv:1410.8494 [hep-lat]}}.

\bibitem{Kogut:2014kla}
J.~B. Kogut and D.~K. Sinclair, ``{Thermodynamics of lattice QCD with 3
  flavours of colour-sextet quarks II. N\_t=6 and N\_t=8}'',
  \href{http://dx.doi.org/10.1103/PhysRevD.90.014506}{{\em Phys. Rev.} {\bf
  D90} (2014) no.~1, 014506},
\href{http://arxiv.org/abs/1406.1524}{{\tt arXiv:1406.1524 [hep-lat]}}.

\bibitem{Kogut:2015zta}
J.~B. Kogut and D.~K. Sinclair, ``{The chiral phase transition for lattice QCD
  with 2 colour-sextet quarks}'',
\href{http://arxiv.org/abs/1507.00375}{{\tt arXiv:1507.00375 [hep-lat]}}.

\bibitem{Fodor:2015zna}
Z.~Fodor, K.~Holland, J.~Kuti, S.~Mondal, D.~Nogradi, and C.~H. Wong, ``{The
  running coupling of the minimal sextet composite Higgs model}'',
\href{http://arxiv.org/abs/1506.06599}{{\tt arXiv:1506.06599 [hep-lat]}}.

\bibitem{Cheng:2014jba}
A.~Cheng, A.~Hasenfratz, Y.~Liu, G.~Petropoulos, and D.~Schaich, ``{Improving
  the continuum limit of gradient flow step scaling}'',
  \href{http://dx.doi.org/10.1007/JHEP05(2014)137}{{\em JHEP} {\bf 1405} (2014)
   137}, \href{http://arxiv.org/abs/1404.0984}{{\tt arXiv:1404.0984}}.

\bibitem{Narayanan:2006rf}
R.~Narayanan and H.~Neuberger, ``{Infinite N phase transitions in continuum
  Wilson loop operators}'',
  \href{http://dx.doi.org/10.1088/1126-6708/2006/03/064}{{\em JHEP} {\bf 0603}
  (2006)  064}, \href{http://arxiv.org/abs/hep-th/0601210}{{\tt
  hep-th/0601210}}.

\bibitem{Luscher:2009eq}
M.~Luscher, ``{Trivializing maps, the Wilson flow and the HMC algorithm}'',
  \href{http://dx.doi.org/10.1007/s00220-009-0953-7}{{\em Commun. Math. Phys.}
  {\bf 293} (2010)  899--919}, \href{http://arxiv.org/abs/0907.5491}{{\tt
  arXiv:0907.5491}}.

\bibitem{Luscher:2013vga}
M.~Luscher, ``{Future applications of the Yang-Mills gradient flow in lattice
  QCD}'', \href{http://pos.sissa.it/archive/conferences/187/016/LATTICE
  2013_016.pdf}{{\em PoS} {\bf LATTICE 2013} (2013)  016},
  \href{http://arxiv.org/abs/1308.5598}{{\tt arXiv:1308.5598}}.

\bibitem{Ramos:2015dla}
A.~Ramos,
  \href{http://inspirehep.net/record/1373921/files/arXiv:1506.00118.pdf}{``{The
  Yang-Mills gradient flow and renormalization}'',} in {\em {Proceedings, 32nd
  International Symposium on Lattice Field Theory (Lattice 2014)}}.
\newblock 2015.
\newblock
\href{http://arxiv.org/abs/1506.00118}{{\tt arXiv:1506.00118 [hep-lat]}}.
\newblock

\bibitem{Luscher:2010iy}
M.~Luscher, ``{Properties and uses of the Wilson flow in lattice QCD}'',
  \href{http://dx.doi.org/10.1007/JHEP08(2010)071}{{\em JHEP} {\bf 1008} (2010)
   071}, \href{http://arxiv.org/abs/1006.4518}{{\tt arXiv:1006.4518}}.

\bibitem{Hasenfratz:2014rna}
A.~Hasenfratz, D.~Schaich, and A.~Veernala, ``{Nonperturbative $\beta$ function
  of eight-flavor SU(3) gauge theory}'',
  \href{http://dx.doi.org/10.1007/JHEP06(2015)143}{{\em JHEP} {\bf 06} (2015)
  143},
\href{http://arxiv.org/abs/1410.5886}{{\tt arXiv:1410.5886 [hep-lat]}}.

\bibitem{Hasenfratz:2015xpa}
A.~Hasenfratz, ``{Improved gradient flow for step scaling function and scale
  setting}'', {\em PoS} {\bf LATTICE2014} (2015)  257,
\href{http://arxiv.org/abs/1501.07848}{{\tt arXiv:1501.07848 [hep-lat]}}.

\bibitem{Fodor:2012td}
Z.~Fodor, K.~Holland, J.~Kuti, D.~Nogradi, and C.~H. Wong, ``{The Yang-Mills
  gradient flow in finite volume}'',
  \href{http://dx.doi.org/10.1007/JHEP11(2012)007}{{\em JHEP} {\bf 1211} (2012)
   007}, \href{http://arxiv.org/abs/1208.1051}{{\tt arXiv:1208.1051}}.

\bibitem{Fodor:2012qh}
Z.~Fodor, K.~Holland, J.~Kuti, D.~Nogradi, and C.~H. Wong, ``{The gradient flow
  running coupling scheme}'',
  \href{http://pos.sissa.it/archive/conferences/164/050/Lattice
  2012_050.pdf}{{\em PoS} {\bf Lattice 2012} (2012)  050},
  \href{http://arxiv.org/abs/1211.3247}{{\tt arXiv:1211.3247}}.

\bibitem{Fritzsch:2013je}
P.~Fritzsch and A.~Ramos, ``{The gradient flow coupling in the Schr{\"o}dinger
  Functional}'', \href{http://dx.doi.org/10.1007/JHEP10(2013)008}{{\em JHEP}
  {\bf 1310} (2013)  008}, \href{http://arxiv.org/abs/1301.4388}{{\tt
  arXiv:1301.4388}}.

\bibitem{Fodor:2014cpa}
Z.~Fodor, K.~Holland, J.~Kuti, S.~Mondal, D.~Nogradi, and C.~H. Wong, ``{The
  lattice gradient flow at tree-level and its improvement}'',
  \href{http://dx.doi.org/10.1007/JHEP09(2014)018}{{\em JHEP} {\bf 1409} (2014)
   018}, \href{http://arxiv.org/abs/1406.0827}{{\tt arXiv:1406.0827}}.

\bibitem{DeGrand:2012qa}
T.~DeGrand, Y.~Shamir, and B.~Svetitsky, ``{SU(4) lattice gauge theory with
  decuplet fermions: Schrodinger functional analysis}'',
  \href{http://dx.doi.org/10.1103/PhysRevD.85.074506}{{\em Phys. Rev.} {\bf
  D85} (2012)  074506},
\href{http://arxiv.org/abs/1202.2675}{{\tt arXiv:1202.2675 [hep-lat]}}.

\bibitem{DeGrand:2014rwa}
T.~DeGrand, Y.~Shamir, and B.~Svetitsky, ``{Suppressing dislocations in
  normalized hypercubic smearing}'',
  \href{http://dx.doi.org/10.1103/PhysRevD.90.054501}{{\em Phys.Rev.} {\bf D90}
  (2014) no.~5, 054501},
\href{http://arxiv.org/abs/1407.4201}{{\tt arXiv:1407.4201 [hep-lat]}}.

\bibitem{Cheng:2011ic}
A.~Cheng, A.~Hasenfratz, and D.~Schaich, ``{Novel phase in SU(3) lattice gauge
  theory with 12 light fermions}'',
  \href{http://dx.doi.org/10.1103/PhysRevD.85.094509}{{\em Phys. Rev.} {\bf
  D85} (2012)  094509}, \href{http://arxiv.org/abs/1111.2317}{{\tt
  arXiv:1111.2317}}.

\bibitem{Cheng:2013xha}
A.~Cheng, A.~Hasenfratz, Y.~Liu, G.~Petropoulos, and D.~Schaich, ``{Finite size
  scaling of conformal theories in the presence of a near-marginal operator}'',
  \href{http://dx.doi.org/10.1103/PhysRevD.90.014509}{{\em Phys. Rev.} {\bf
  D90} (2014)  014509}, \href{http://arxiv.org/abs/1401.0195}{{\tt
  arXiv:1401.0195}}.

\bibitem{Ishikawa:2013tua}
K.~I. Ishikawa, Y.~Iwasaki, Y.~Nakayama, and T.~Yoshie, ``{Global Structure of
  Conformal Theories in the SU(3) Gauge Theory}'',
  \href{http://dx.doi.org/10.1103/PhysRevD.89.114503}{{\em Phys. Rev.} {\bf
  D89} (2014) no.~11, 114503},
\href{http://arxiv.org/abs/1310.5049}{{\tt arXiv:1310.5049 [hep-lat]}}.

\bibitem{Ishikawa:2013wf}
K.-I. Ishikawa, Y.~Iwasaki, Y.~Nakayama, and T.~Yoshie, ``{Conformal theories
  with an infrared cutoff}'',
  \href{http://dx.doi.org/10.1103/PhysRevD.87.071503}{{\em Phys.Rev.} {\bf D87}
  (2013) no.~7, 071503},
\href{http://arxiv.org/abs/1301.4785}{{\tt arXiv:1301.4785 [hep-lat]}}.

\bibitem{Aoki:1983qi}
S.~Aoki, ``{New Phase Structure for Lattice QCD with Wilson Fermions}'',
\href{http://dx.doi.org/10.1103/PhysRevD.30.2653}{{\em Phys. Rev.} {\bf D30}
  (1984)  2653}.

\bibitem{Sharpe:1998xm}
S.~R. Sharpe and R.~L. Singleton, Jr, ``{Spontaneous flavor and parity breaking
  with Wilson fermions}'',
  \href{http://dx.doi.org/10.1103/PhysRevD.58.074501}{{\em Phys. Rev.} {\bf
  D58} (1998)  074501},
\href{http://arxiv.org/abs/hep-lat/9804028}{{\tt arXiv:hep-lat/9804028
  [hep-lat]}}.

\bibitem{Lin:2014fxa}
C.~J.~D. Lin, K.~Ogawa, H.~Ohki, A.~Ramos, and E.~Shintani, ``{SU(3) gauge
  theory with 12 flavours in a twisted box}'', {\em PoS} {\bf LATTICE2014}
  (2014)  259,
\href{http://arxiv.org/abs/1410.8824}{{\tt arXiv:1410.8824 [hep-lat]}}.

\bibitem{Calabrese:2002bm}
P.~Calabrese, A.~Pelissetto, and E.~Vicari, ``{Multicritical phenomena in
  O(n(1)) + O(n(2)) symmetric theories}'',
  \href{http://dx.doi.org/10.1103/PhysRevB.67.054505}{{\em Phys. Rev.} {\bf
  B67} (2003)  054505},
\href{http://arxiv.org/abs/cond-mat/0209580}{{\tt arXiv:cond-mat/0209580
  [cond-mat]}}.

\end{thebibliography}\endgroup
\end{document}